\documentclass
[aps,prb,amsfonts,amssymb,onecolumn,amsmath,floatfix,notitlepage,longbibliography,superscriptaddress]{revtex4-2}
\usepackage{graphics}
\usepackage{natbib}
\usepackage{comment}
\usepackage{graphicx}
\usepackage{bm}
\usepackage{amsmath}
\usepackage{extarrows}
\usepackage{amsfonts}
\usepackage{amssymb}
\usepackage{xcolor}
\usepackage{tabularx}
\usepackage{multirow}
\usepackage{makecell}
\usepackage{subfigure}
\usepackage[colorlinks=true, pdfstartview=FitV,  linkcolor=blue, citecolor=blue, urlcolor=blue]{hyperref}
\usepackage{hypcap}
\usepackage{braket}
\usepackage{commath}%
\usepackage[percent]{overpic}
\providecommand{\U}[1]{\protect\rule{.1in}{.1in}}
\begin{document}

\title{Quantum Algorithm Software for Condensed Matter Physics}

\author{T. Farajollahpour}
\affiliation{Department of Physics, Norwegian University of Science and Technology (NTNU), NO-7491 Trondheim, Norway}
\affiliation{Department of Physics, Brock University, St. Catharines, Ontario L2S 3A1, Canada}

\begin{abstract}
Realizing the promise of quantum computation for condensed matter many-body
problems depends as much on software as on hardware, yet the area is reviewed far
more often than it is quantified. We address this gap by pairing a focused survey
of quantum algorithm software for condensed matter physics with a compact, fully
reproducible benchmark suite that turns qualitative claims into concrete numbers.
Each algorithm family, namely the variational quantum eigensolver (VQE), quantum
phase estimation (QPE), quantum annealing and the quantum approximate optimization
algorithm (QAOA), and quantum machine learning (QML), is demonstrated on a
canonical lattice model and validated against an independent classical reference,
from exact diagonalization and the Bethe ansatz to matrix-product-state DMRG.
Within this suite we quantify two issues usually treated only qualitatively.
Mapping the Fermi-Hubbard model to qubits under the Jordan-Wigner and Bravyi-Kitaev
encodings, we tabulate qubit counts, operator weights, and gate costs and expose a
geometry-dependent trade-off between the two. Simulating the circuits under a
depolarizing noise model, we show that zero-noise extrapolation restores
ground-state energies and optimization quality across the noise range. Around these
results we review the algorithms as applied to strongly correlated systems,
topological phases, and quantum magnetism, together with the leading software
development kits (Qiskit, Cirq, PennyLane, and Q\#) and the classical and
tensor-network methods against which quantum approaches must be benchmarked. All
circuits, seeds, and data are released so the benchmarks can be reproduced and
extended. We argue that standardized, reproducible benchmarks of this kind are
essential to gauge progress and identify genuine quantum advantage in condensed
matter physics.
\end{abstract}

\maketitle
\tableofcontents



\section{The Quantum Frontier in Condensed Matter Physics}

The study of condensed matter physics, which seeks to understand the macroscopic and microscopic physical properties of matter, presents some of the most computationally challenging problems in modern science. Many of these challenges stem from the complex quantum mechanical interactions of a vast number of constituent particles, such as electrons in a solid~\cite{altland2010condensed}. Classical computational methods, despite their sophistication, encounter fundamental limitations when attempting to simulate these quantum many-body systems accurately. The core difficulty lies in the exponential scaling of the Hilbert space with the number of particles, a phenomenon often referred to as the "curse of dimensionality"~\cite{bellman1957dynamic}. This exponential growth makes exact classical simulations intractable for even moderately sized systems, hindering progress in understanding and predicting the behavior of novel materials and quantum phenomena.

Quantum computers, by directly leveraging the principles of quantum mechanics like superposition and entanglement, offer a promising pathway to overcome these classical limitations~\cite{mermin2007quantum,nielsen2010quantum,wong2022introduction}. Condensed matter physics is a particularly fertile ground for quantum algorithms, with numerous problems poised to benefit from quantum computational approaches~\cite{dalzell2310quantum}. Key areas include the study of \text{strongly correlated systems}, where electron-electron interactions dominate and give rise to exotic behaviors such as high-temperature superconductivity and complex magnetic ordering~\cite{dalzell2310quantum}. Other significant applications involve the characterization of \text{topological phases of matter}, which possess robust properties determined by global topology rather than local order parameters and are relevant for fault-tolerant quantum computing~\cite{Pollmann2024}. Understanding \text{quantum magnetism}, the collective behavior of quantum spins in materials, and simulating various \text{lattice models} like the Fermi-Hubbard, Heisenberg, and Ising models, which serve as canonical representations of fundamental condensed matter phenomena, are also prime targets for quantum algorithms~\cite{DiVincenzo2005,Raussendorf2007,Pastawski2023}.

The successful application of quantum computation to these problems hinges critically on the development of sophisticated quantum algorithm software. This software acts as the essential bridge, translating abstract theoretical algorithms into concrete instructions that can be executed on either existing quantum hardware or classical simulators~\cite{Nunez2025productive}. It encompasses a range of tools from high-level programming languages and circuit construction libraries to compilers, optimizers, and interfaces with physical quantum processing units (QPUs). The development of such software is not merely an engineering task of implementing pre-defined algorithms. Rather, it is an active and dynamic research area in itself. This field is characterized by a co-evolutionary relationship with both quantum hardware development and theoretical algorithmic advancements. The constraints and capabilities of current Noisy Intermediate-Scale Quantum (NISQ) devices~\cite{Preskill2018}, for example, heavily influence algorithm design and software optimization strategies, fostering a "co-design" paradigm where algorithms, software, and hardware are developed in tandem to maximize performance and practical utility~\cite{Mueller2023,Aharonov2025}.

Surveys of this landscape are now plentiful, but they tend to describe algorithms and software qualitatively rather than measure them, which makes it difficult to compare encodings, gauge the cost of noise, or track progress against a fixed reference. The present work is built around closing that gap. Its central contribution is a compact and fully reproducible benchmark suite in which each principal algorithm family is demonstrated on a canonical lattice model and validated against an independent classical reference, with all circuits, random seeds, and generated data released so that the results can be reproduced and extended. Two questions that the literature usually leaves at the level of intuition are made quantitative. We map the Fermi-Hubbard model to qubits under the Jordan-Wigner and Bravyi-Kitaev encodings and tabulate the resulting qubit counts, operator weights, and gate costs, exposing a geometry-dependent trade-off between the two, and we simulate the variational circuits under a depolarizing noise model to show how zero-noise extrapolation recovers ground-state energies and optimization quality across a realistic noise range.

Around these results we provide the survey that gives them context, covering the fundamental quantum algorithms employed and developed for condensed matter physics, the leading software development kits (SDKs) and libraries, the classical and tensor-network methods against which quantum approaches must be benchmarked, and the current challenges of hardware limitations and algorithmic scalability together with the future trajectories of the field. The aim is a reference that serves both as a map of the area and as a concrete, reusable yardstick for measuring progress toward genuine quantum advantage in condensed matter systems.

\section{Fundamental Quantum Algorithms for Condensed Matter Physics}
\label{sec:algorithms}
A diverse suite of quantum algorithms is being developed and refined to tackle the
complex problems inherent in condensed matter systems. These algorithms leverage quantum
mechanical principles to offer potential advantages over classical computational methods.
Key approaches include the Variational Quantum Eigensolver (VQE) for ground state problems~\cite{Peruzzo2014variational},
Quantum Phase Estimation (QPE) for spectral properties~\cite{Kitaev1995quantum}, Quantum Annealing (QA)~\cite{Apolloni1990}
and the Quantum Approximate Optimization Algorithm (QAOA) for optimization tasks mappable
to condensed matter Hamiltonians~\cite{BLEKOS20241}, and various forms of Quantum Machine Learning (QML)
for tasks like phase classification and materials discovery~\cite{Biamonte2017}. Additionally, specialized
quantum simulations target specific models like the Fermi-Hubbard~\cite{Stanisic2022observing} and Heisenberg models~\cite{Barry1976},
and tensor network methods provide both classical benchmarks and inspiration for quantum algorithms.

\begin{figure*}[t]
\centering
\includegraphics[width=\linewidth]{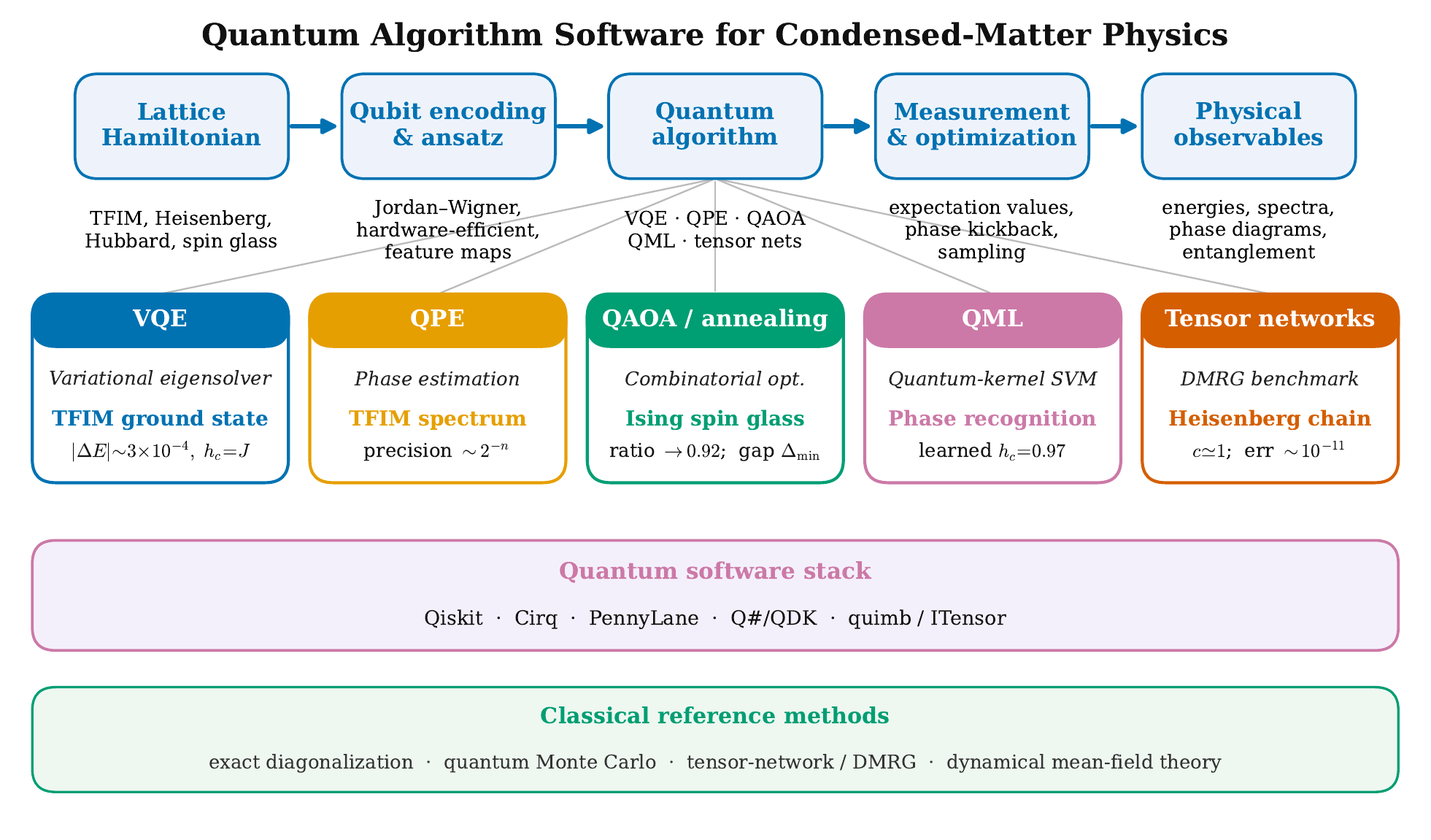}
\caption{Overview of the quantum-algorithm workflow for condensed-matter
physics surveyed in this review. A lattice Hamiltonian is mapped onto qubits
together with a variational or time-evolution ansatz, processed by a quantum
algorithm, and read out through measurement and (where applicable) a classical
optimization loop to yield physical observables. The five algorithm families
discussed here, the variational quantum eigensolver (VQE), quantum phase
estimation (QPE), quantum annealing and the quantum approximate optimization
algorithm (QAOA), quantum machine learning (QML), and tensor-network
methods, are each illustrated in this section on a canonical model, with the
quantitative outcome of the corresponding worked example
(Figs.~\ref{fig:vqe}-\ref{fig:tn}) indicated. All are supported by the quantum
software stack (Qiskit, Cirq, PennyLane, Q\#/QDK, \texttt{quimb}/ITensor) and
validated against classical reference methods.}
\label{fig:schematic}
\end{figure*}

\subsection{Variational Quantum Eigensolver (VQE)}

The Variational Quantum Eigensolver (VQE) is a hybrid quantum-classical algorithm designed primarily to find the ground state energy of a given Hamiltonian, a central task in many condensed matter problems~\cite{Peruzzo2014variational,Li2025,Novak2025}. The VQE operates by preparing a trial quantum state using a parameterized quantum circuit, known as an ansatz, on a quantum processor. The energy of this trial state with respect to the target Hamiltonian is then measured. This energy value is fed to a classical optimization algorithm, which iteratively adjusts the parameters of the quantum circuit to minimize the measured energy, thereby converging towards an approximation of the ground state energy according to the variational principle~\cite{Li2025}.

VQE has found numerous applications in condensed matter physics. It is employed for simulating molecular Hamiltonians, which can be relevant for understanding material properties at a fundamental level~\cite{Peruzzo2014variational,Kandala2017,Tilly2022variational}. A significant area of application is the study of quantum phase transitions, where VQE can help map out the energy landscape as system parameters are varied~\cite{Tilly2022variational,Lively2024,Kirmani2024}. It is also used to explore exotic states of matter~\cite{Kandala2017} and to determine the ground state properties of important lattice models, most notably the Fermi-Hubbard model~\cite{Arute2020,Google2020,Stanisic2022observing}, which is crucial for understanding phenomena like high-temperature superconductivity. Recent advancements include the development of techniques like the Knowledge Distillation Inspired VQE (KD-VQE), which has shown improved convergence behavior when applied to the Fermi-Hubbard model by utilizing a collection of trial wavefunctions governed by a Boltzmann distribution with a virtual temperature. VQE is also being benchmarked for solving partial differential equations (PDEs) that can describe physical phenomena, such as the advection-diffusion equation, by encoding each time step into a ground-state problem~\cite{lubasch2020,Kyriienko2021,Sun2021}.

The primary strength of VQE lies in its potential utility in the NISQ era. Compared to algorithms like Quantum Phase Estimation, VQE typically requires shallower quantum circuits, making it more resilient to the noise and limited coherence times of current quantum hardware~\cite{Peruzzo2014variational,Mcclean2016theory,Preskill2018,Bharti2022noisy}. However, VQE is not without its challenges. The design of an effective ansatz that can accurately represent the true ground state while remaining trainable is a critical hurdle. Many VQE implementations suffer from the "barren plateau" phenomenon, where gradients vanish exponentially with system size, making optimization intractable; for sufficiently deep or unstructured ansatze the gradient variance is exponentially small in the qubit number, $\mathrm{Var}[\partial_{\theta_k}\langle H\rangle]\in\mathcal{O}(b^{-L})$ with $b>1$. The energy itself is estimated by expanding the Hamiltonian into Pauli strings, $H=\sum_a c_a P_a$, and averaging $\langle H\rangle=\sum_a c_a\braket{\psi(\bm\theta)|P_a|\psi(\bm\theta)}$, so that the measurement budget grows with the number of terms; this overhead, together with the choice of classical optimizer, poses significant practical difficulties~\cite{Grimsley2019,Tang2021qubit,Mcclean2018barren,Cerezo2021cost,Gonthier2022}.

The performance of VQE is intimately tied to the specific problem Hamiltonian and the chosen ansatz. Generic, hardware-efficient ansatze often struggle to capture the complex correlations present in many condensed matter ground states. This necessitates the development of problem-specific or "Hamiltonian-aware" ansatze. Adaptive ansatz construction strategies, where the structure of the circuit is grown iteratively based on some physical or information-theoretic criterion, represent a promising direction. As an example, the Overlap-ADAPT-VQE algorithm iteratively generates a compact approximation of a target wave function by maximizing overlap at each step~\cite{Feniou2023overlap}. Furthermore, techniques like Quantum Architecture Search (QAS) are being explored to automate the design of efficient parameterized quantum circuits~\cite{Zhang2022quantum,Du2022quantum,Chivilikhin2020mog}. These developments highlight a crucial aspect of VQE, namely that its success is not solely dependent on hardware improvements but also on the sophisticated co-design of the quantum circuit and the classical optimization loop, tailored to the physics of the system under investigation. Future VQE software will likely need to incorporate advanced tools for automated ansatz generation, optimization landscape analysis, and intelligent resource allocation to navigate these challenges effectively.

\begin{figure*}[t]
\centering
\includegraphics[width=\linewidth]{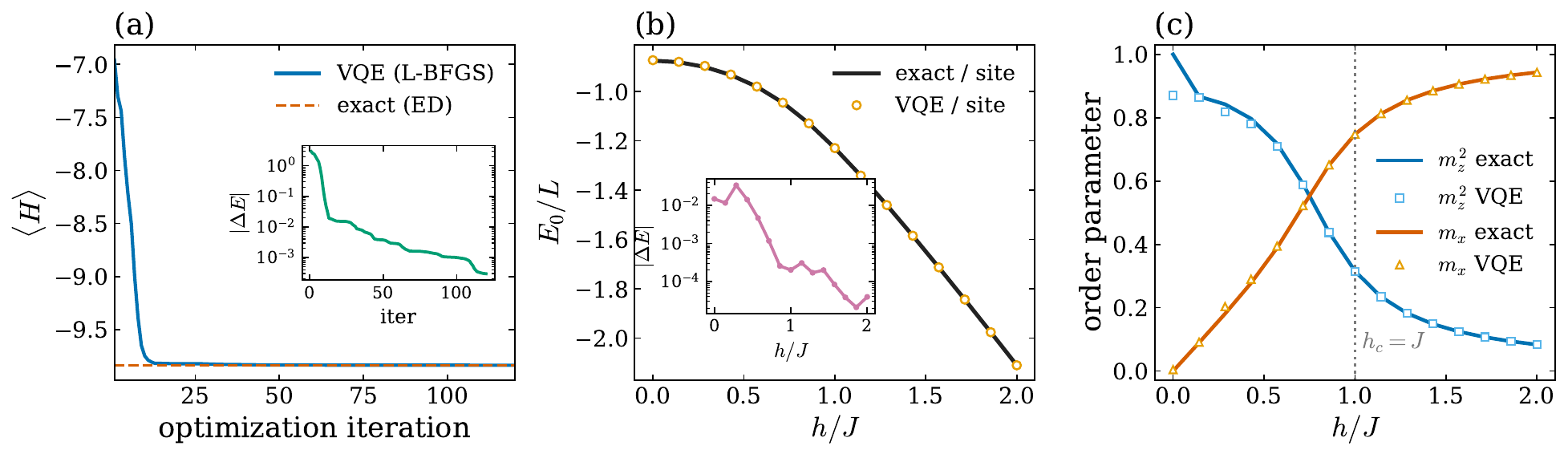}
\caption{VQE for the $L=8$ transverse-field Ising chain,
Eq.~\eqref{eq:tfim}. (a)~Energy convergence at criticality ($h=J$). The inset
shows the optimization error $|\Delta E|$ on a logarithmic scale. (b)~Ground
-state energy per site across the transverse-field sweep, compared with exact
diagonalization, with the inset showing the error versus $h/J$. (c)~Squared longitudinal order
parameter $m_z^2$ and transverse magnetization $m_x$, locating the quantum
phase transition at $h_c=J$ (dotted line). Symbols are VQE and solid curves are
exact.}
\label{fig:vqe}
\end{figure*}

To make the discussion above concrete, we apply VQE to the one-dimensional
transverse-field Ising model (TFIM),
\begin{equation}
H_{\rm TFIM}=-J\sum_{i}Z_iZ_{i+1}-h\sum_i X_i ,
\label{eq:tfim}
\end{equation}
a textbook host of a quantum phase transition at $h_c=J$. The variational
principle bounds the energy of any trial state from below by the ground-state
energy,
\begin{equation}
E(\bm{\theta})=\frac{\braket{\psi(\bm{\theta})|H_{\rm TFIM}|\psi(\bm{\theta})}}
{\braket{\psi(\bm{\theta})|\psi(\bm{\theta})}}\ \ge\ E_0 ,
\label{eq:varprinciple}
\end{equation}
and we minimize it over a Hamiltonian-variational ansatz of depth $P$,
\begin{equation}
\ket{\psi(\bm{\theta})}=\prod_{l=1}^{P}e^{\,i\beta_l\sum_i X_i}\,
e^{\,i\gamma_l\sum_i Z_iZ_{i+1}}\ket{+}^{\otimes L},
\label{eq:hva}
\end{equation}
with one variational pair $(\gamma_l,\beta_l)$ per layer. On $L=8$ spins at
depth $P=8$, optimized with the L-BFGS method and adjoint gradients
(\texttt{PennyLane}\,/\,\texttt{lightning}), the energy converges at criticality
to the exact ground state to $|\Delta E|\!\sim\!3\times10^{-4}$
[Fig.~\ref{fig:vqe}(a)]. Sweeping the transverse field reproduces the exact
ground-state energy across the phase diagram [Fig.~\ref{fig:vqe}(b)], while the
longitudinal and transverse order parameters
\begin{equation}
m_x=\frac1L\sum_i\braket{X_i},\qquad
m_z^2=\frac1{L^2}\sum_{ij}\braket{Z_iZ_j},
\label{eq:orderparams}
\end{equation}
resolve the quantum critical point at $h_c=J$ [Fig.~\ref{fig:vqe}(c)]. The
residual error grows in the deep ordered phase
($h\!\to\!0$), a direct manifestation of the trainability difficulty associated
with the near-degenerate symmetry-broken ground state, and an honest reflection
of the ansatz and optimizer limitations discussed above.

\subsection{Quantum Phase Estimation (QPE)}
Quantum Phase Estimation (QPE) is a cornerstone quantum algorithm designed to determine the eigenvalues of a unitary operator, which, for a given Hamiltonian $H$, corresponds to finding its energy eigenvalues if the input state is an eigenstate~\cite{nielsen2010quantum,Kitaev1995quantum,Cleve1998quantum}. The algorithm typically involves preparing an auxiliary register of qubits in a superposition state, applying controlled unitary operations (powers of the evolution operator $e^{-iHt}$) conditioned on these auxiliary qubits, and then performing an inverse Quantum Fourier Transform (QFT) on the auxiliary register. Measuring this register yields a binary representation of the phase, which directly relates to the eigenvalue~\cite{nielsen2010quantum,Cleve1998quantum}.

In condensed matter physics, QPE is primarily targeted at determining the ground and excited state energies of various Hamiltonians, providing fundamental insights into material properties and molecular interactions. Beyond static energy calculations, QPE can be a subroutine in algorithms for calculating spectral properties and dynamic response functions, which are crucial for understanding how materials interact with external probes and for predicting spectroscopic signatures~\cite{Aspuru2005simulated,Babbush2015chemical,Poulin2009quantum,Ortiz2001quantum}.

The main strength of QPE lies in its potential to achieve an exponential \textit{speedup} over classical methods for finding eigenvalues with high precision, provided certain conditions are met. However, the algorithm faces significant challenges, particularly in the NISQ era. QPE typically requires deep quantum circuits due to the controlled unitary operations and the QFT, making it highly susceptible to errors on current noisy hardware. Standard QPE applies $\sum_{j=0}^{n-1}2^{j}=2^{n}-1$ controlled-unitary calls to resolve $n$ bits and returns the best $n$-bit estimate with probability at least $4/\pi^2\approx0.41$~\cite{nielsen2010quantum}, so improving precision is exponentially costly in circuit depth. Furthermore, its success probability and accuracy depend critically on the quality of the initial state preparation, and the input state must have a significant overlap with the target eigenstate~\cite{nielsen2010quantum,Preskill2018,Bharti2022noisy,Tilly2022variational}. Preparing such high-fidelity initial states for complex condensed matter systems is a non-trivial task in itself.

Recognizing these limitations, recent research has focused on
developing more hardware-friendly variants of QPE. A notable
advancement is the development of "control-free" QPE methods.
These approaches aim to eliminate the need for costly controlled
time evolution operations by leveraging techniques from classical
signal processing, such as phase retrieval algorithms~\cite{Somma2019quantum,Dong2021ground}.
By measuring properties of the uncontrolled time evolution of the system (e.g., expectation values of certain operators at different times) and then classically post-processing this data, it is possible to reconstruct the spectral information. Such methods have been numerically investigated for models like the Fermi-Hubbard model and show promise for reducing circuit depth and simplifying implementation on near-term devices. Another important development is the Generalized Quantum Phase Estimation (GQPE) framework, which extends QPE to multi-variate expectations, enabling the calculation of higher-order correlation functions and, consequently, nonlinear response properties relevant for advanced spectroscopy~\cite{Loaiza2024nonlinear}. These innovations signal a significant shift in QPE research, moving towards algorithms that are more robust to the imperfections of current and near-future quantum hardware. This evolution is critical for accelerating the application of phase estimation techniques to challenging problems in condensed matter physics, rather than solely relying on the eventual advent of large-scale, fault-tolerant quantum computers.

\begin{figure*}[t]
\centering
\includegraphics[width=0.86\linewidth]{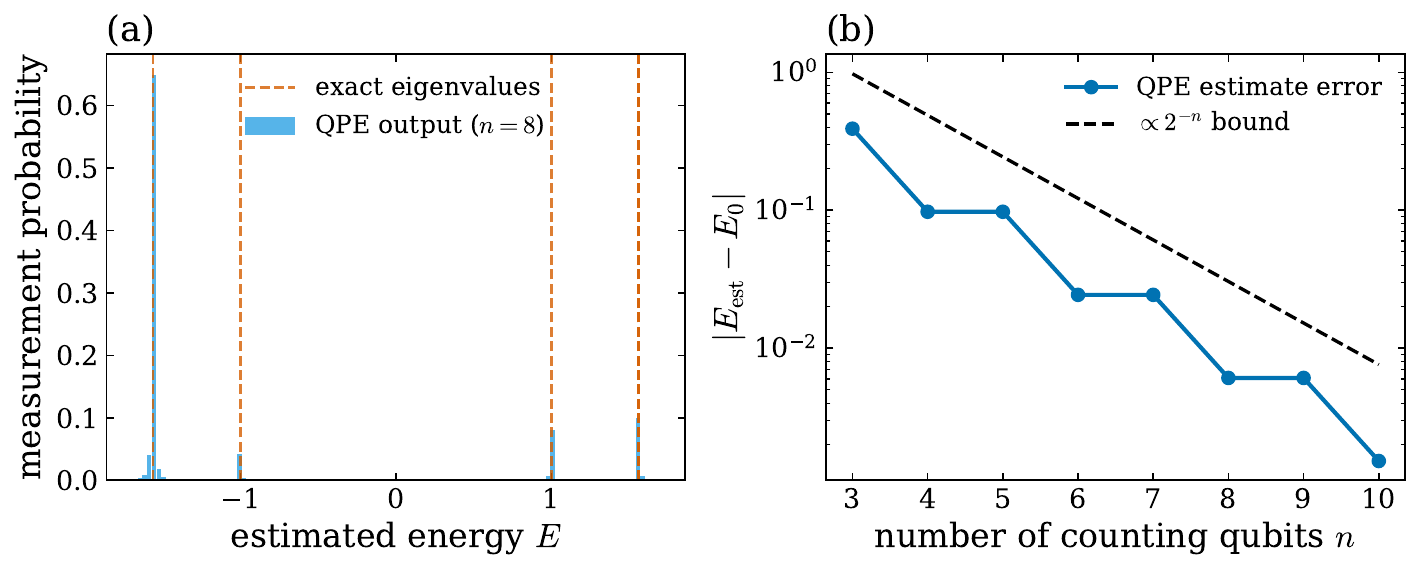}
\caption{QPE for a two-site transverse-field Ising model. (a)~Counting-register
measurement probability for a generic input state and $n=8$ counting qubits, where
peaks coincide with the exact eigenvalues (dashed), with heights given by the
eigenstate overlaps. (b)~Ground-state energy error versus the number of
counting qubits $n$, following the $\propto 2^{-n}$ bound (dashed).}
\label{fig:qpe}
\end{figure*}

To illustrate phase estimation we map the spectrum of a two-site TFIM into the
phase register of an idealized QPE circuit. Writing the evolution operator as
$U=e^{-iH\tau}$, each eigenstate carries a phase fixed by its energy,
\begin{equation}
U\ket{\phi_k}=e^{2\pi i\phi_k}\ket{\phi_k},\qquad
\phi_k=-\frac{E_k\tau}{2\pi}\ (\mathrm{mod}\ 1),
\label{eq:qpephase}
\end{equation}
with $\tau$ chosen so that every phase lies in $[0,1)$. We evaluate the
counting-register distribution exactly through its Dirichlet (Fej\'er) kernel:
for a true phase $\phi$ the probability of reading the $n$-bit outcome $y$ is
\begin{equation}
P(y)=\frac{1}{2^{2n}}\,
\frac{\sin^2\!\big(2^{n}\pi\,\delta\big)}{\sin^2\!\big(\pi\,\delta\big)},
\qquad \delta=\phi-\frac{y}{2^{n}}.
\label{eq:fejer}
\end{equation}
For a generic input state with nonzero overlap
on every eigenstate, the simulated measurement histogram reproduces the full
set of eigenvalues, with peak weights set by the squared overlaps
$|\!\braket{\phi_k|\psi_{\rm in}}\!|^2$ [Fig.~\ref{fig:qpe}(a)]. Preparing the
exact ground state instead, the energy estimate converges to the true value as
the register is enlarged, gaining roughly one bit of precision per added
counting qubit, $|E_{\rm est}-E_0|\sim 2^{-n}$ [Fig.~\ref{fig:qpe}(b)], the
Heisenberg-limited scaling that underlies the asymptotic advantage of QPE, here
displayed alongside its practical dependence on initial-state overlap.

\subsection{Quantum Annealing (QA) and Quantum Approximate Optimization Algorithm (QAOA)}

Quantum Annealing (QA) is a metaheuristic approach to finding the ground state of a problem Hamiltonian, typically an Ising Hamiltonian, by leveraging quantum tunneling~\cite{Kadowaki1998quantum,Farhi2001quantum,Das2008colloquium,Hauke2020}. The process involves initializing a system of qubits in the easy-to-prepare ground state of a driver Hamiltonian (often a transverse field). The Hamiltonian of the system is then slowly (adiabatically) evolved from this driver Hamiltonian to the problem Hamiltonian whose ground state encodes the solution to an optimization problem~\cite{Farhi2001quantum,Das2008colloquium,Hauke2020}. If the evolution is sufficiently slow, the quantum adiabatic theorem guarantees that the system remains in its instantaneous ground state, ideally ending in the ground state of the problem Hamiltonian. The Quantum Approximate Optimization Algorithm (QAOA) is a hybrid quantum-classical algorithm often used for similar optimization problems but implemented on gate-based quantum computers. It involves iteratively applying a parameterized sequence of operators related to the problem and driver Hamiltonians and classically optimizing these parameters~\cite{Farhi2001quantum}.

In condensed matter physics, QA and QAOA are primarily applied to problems that can be mapped to finding the ground state of an Ising-like Hamiltonian. A prominent example is determining the ground states of spin glasses, which are notoriously difficult classical optimization problems due to their frustrated interactions and complex energy landscapes. QA is also used to study the dynamics of annealing protocols themselves and their relation to thermalization and quantum phase transitions~\cite{Bapst2012quantum,Rajak2023quantum,Yarkoni2022quantum}.

A key strength of QA is the relative maturity of dedicated quantum annealing hardware, such as those developed by D-Wave Systems, which feature thousands of qubits designed for such optimization tasks~\cite{Dwave,Hauke2020,King2023}. QAOA offers a gate-based alternative that can be explored on universal quantum computers. However, the applicability of QA is largely limited to optimization problems that can be cast in the Ising form. A significant challenge remains in demonstrating a clear and unambiguous quantum advantage over sophisticated classical optimization heuristics for practically relevant problems. Embedding the problem graph onto the often sparse and fixed connectivity of QA hardware can also be a complex task~\cite{Quinton2025}.

Recent advancements in this area are multifaceted. For QA, techniques like Learning-Driven Annealing (LDA) are being developed to mitigate hardware constraints such as short annealing times and control errors by adaptively modifying the problem Hamiltonian based on information learned from previous annealing runs~\cite{Grajciar2024learning}. Experimental results from D-Wave's 3D annealer with up to $5627$ qubits have shown promising scaling for finding exact ground states of $3$D Ising spin glasses, potentially outperforming known classical exact algorithms for certain system sizes~\cite{Smelyanskiy2025computational}. Conversely, classical simulation methods for quantum annealing are also improving. For instance, time-dependent Variational Monte Carlo (t-VMC) using Jastrow-Feenberg wave functions has been shown to efficiently simulate the quantum annealing of spin glasses up to system sizes previously thought intractable for such classical methods, achieving accuracy comparable to quantum processing units (QPUs) with only polynomially scaling classical resources~\cite{Fabiani2024simulating}. This classical progress continuously raises the bar for demonstrating quantum advantage. For QAOA, research focuses on better parameter optimization strategies and understanding its connection to adiabatic evolution, with some studies showing convergence of QAOA angles to universal QA trajectories~\cite{Zhou2020quantum,Fabiani2024simulating}.

The ongoing competition between improving quantum annealing hardware/algorithms and advancing classical simulation techniques for these specific optimization problems highlights a critical aspect of the field, namely that the frontier for ''quantum advantage" is not static. Demonstrating robust quantum advantage requires outperforming the \textit{best} available classical algorithms, including specialized heuristics, not just generic or brute-force classical methods~\cite{Kim2023utility,Begusic2024fast,Tindall2024efficient}. This necessitates continuous, rigorous benchmarking and a co-development strategy where quantum approaches are constantly evaluated against the evolving landscape of classical computation.

\begin{figure*}[t]
\centering
\includegraphics[width=\linewidth]{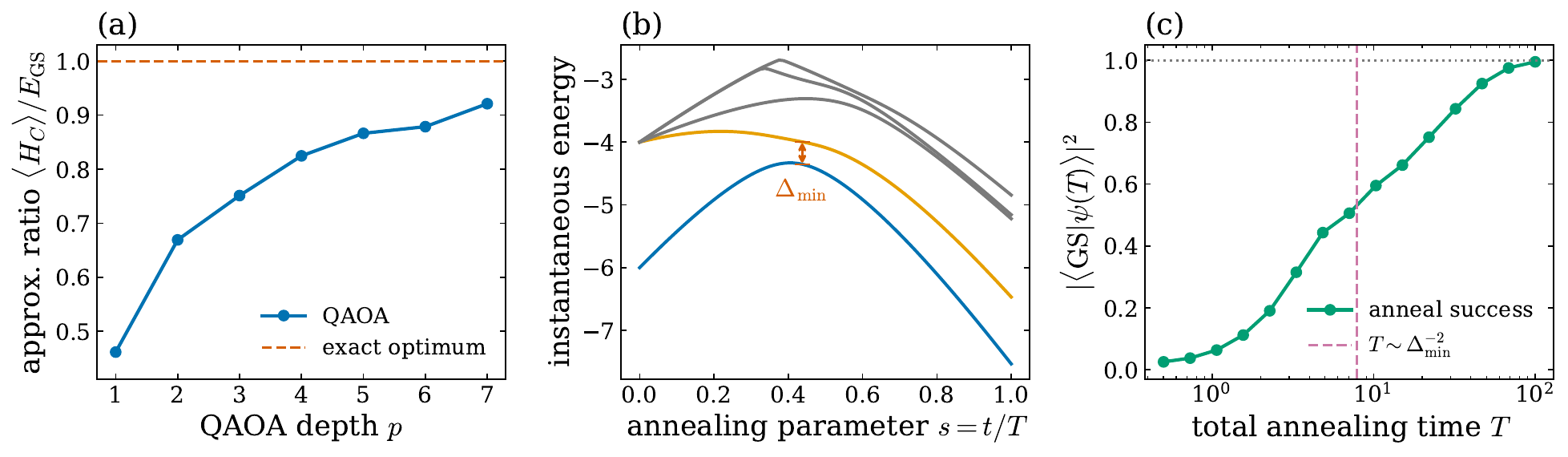}
\caption{Optimization of an $L=6$ Ising spin glass with a unique ground state.
(a)~QAOA approximation ratio versus circuit depth $p$. (b)~Instantaneous
spectrum of the annealing Hamiltonian $H(s)$, where the double arrow marks the
minimum gap $\Delta_{\min}=0.36$. (c)~Adiabatic success probability
$|\!\braket{\mathrm{GS}|\psi(T)}\!|^2$ versus total annealing time $T$, where the
dashed line indicates the $T\sim\Delta_{\min}^{-2}$ scale.}
\label{fig:qaoa}
\end{figure*}
\begin{figure*}[t]
\centering
\includegraphics[width=\linewidth]{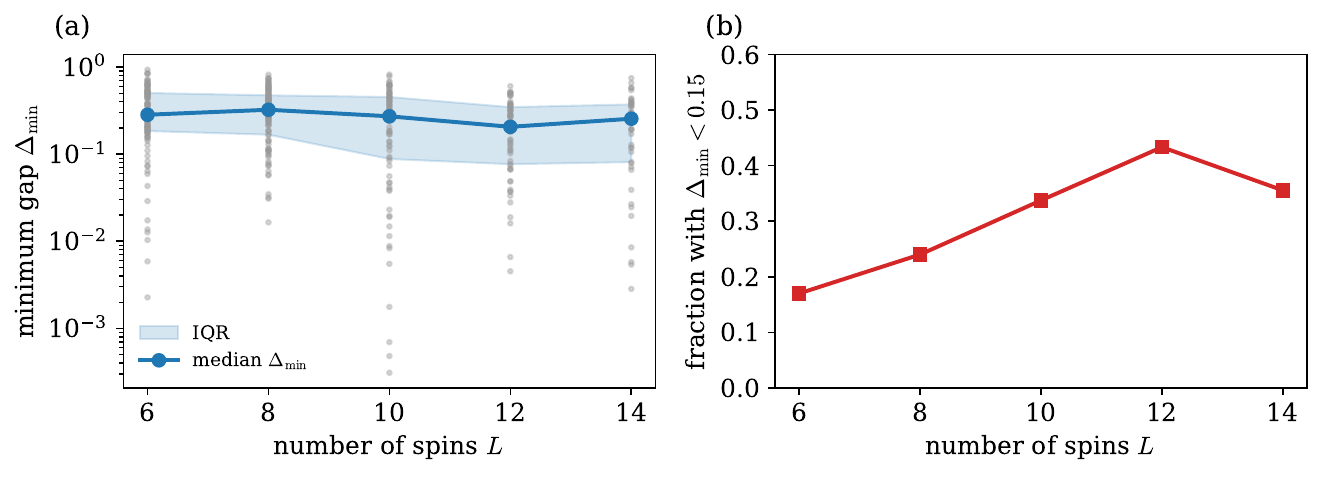}
\caption{Minimum-gap statistics of random frustrated Ising spin glasses (a ring
of $L$ sites plus $L/2$ random chords, $J_{ij}\in\{\pm1\}$,
$b_i\sim\mathcal{N}(0,0.35)$, restricted to unique-ground-state instances)
versus system size, generalizing the single instance of Fig.~\ref{fig:qaoa}.
(a)~Per-instance $\Delta_{\min}$ (grey points), median (blue), and interquartile
range: over the exact-diagonalization range $L\le14$ the median is roughly
stable while the lower tail of the distribution grows, i.e.\ small-gap instances
become increasingly common. (b)~Fraction of ``hard'' instances with
$\Delta_{\min}<0.15$, which rises from $\sim\!0.17$ to $\sim\!0.4$ and saturates.
This is the finite-size precursor of the gap-controlled annealing slowdown
$T\sim\Delta_{\min}^{-2}$; resolving the asymptotic scaling cleanly would require
larger systems and ensembles than ED permits. Statistics use $45$--$100$
instances per size.}
\label{fig:gap_scaling}
\end{figure*}

We benchmark both gate-based QAOA and adiabatic annealing on a frustrated Ising
spin glass of $L=6$ spins with random $\pm1$ couplings and a unique ground
state, $H_C=\sum_{\langle ij\rangle}J_{ij}Z_iZ_j+\sum_i b_iZ_i$. The QAOA prepares
a depth-$p$ alternating state and minimizes the cost expectation over its angles,
\begin{equation}
\ket{\bm\gamma,\bm\beta}=\prod_{l=1}^{p}e^{-i\beta_l H_B}\,e^{-i\gamma_l H_C}
\ket{+}^{\otimes L},\quad H_B=\sum_i X_i,\quad
F_p=\braket{\bm\gamma,\bm\beta|H_C|\bm\gamma,\bm\beta},
\label{eq:qaoa}
\end{equation}
its performance measured by the approximation ratio $r=\langle H_C\rangle/E_{\rm
GS}$. With the variational angles optimized at each depth, QAOA improves
monotonically from $r=0.46$ at $p=1$ to $0.92$ at $p=7$
[Fig.~\ref{fig:qaoa}(a)]. The adiabatic comparison interpolates
$H(s)=(1-s)H_{\rm driver}+sH_C$ with $s=t/T$, where the adiabatic theorem
requires a total evolution time set by the inverse square of the minimum gap,
\begin{equation}
T\ \gg\ \frac{\max_s\lVert\partial_s H(s)\rVert}{\Delta_{\min}^{2}},\qquad
\Delta_{\min}=\min_s\big[E_1(s)-E_0(s)\big].
\label{eq:adiabatic}
\end{equation}
The instantaneous spectrum displays a minimum gap $\Delta_{\min}=0.36$ near
$s\approx0.44$ [Fig.~\ref{fig:qaoa}(b)], and the corresponding annealing success
probability rises from $\sim\!0.03$ for fast schedules to $\sim\!0.99$ once the
total evolution time exceeds the characteristic adiabatic scale
$T\sim\Delta_{\min}^{-2}$ [Fig.~\ref{fig:qaoa}(c)]. The example concretely links
the two optimization paradigms discussed above and exposes the gap-controlled
runtime that governs quantum annealing. Extending the single instance to random
ensembles across a range of sizes, the minimum gap develops a growing tail of
small-gap (hard) instances as $L$ increases [Fig.~\ref{fig:gap_scaling}], the
finite-size precursor of the gap-controlled slowdown that ultimately limits the
approach.

\subsection{Quantum Machine Learning (QML)}

Quantum Machine Learning (QML) represents an interdisciplinary field that explores the intersection of quantum computing and machine learning. This can involve using quantum computers to potentially accelerate or enhance classical machine learning tasks, or applying classical machine learning techniques to analyze data generated from quantum systems or to assist in the design and control of quantum experiments and algorithms~\cite{Biamonte2017,Schuld2015introduction,Dunjko2018machine,Cerezo2022challenges}

In condensed matter physics, QML is emerging as a versatile tool with several promising applications. One key area is the \textit{classification of quantum phases of matter}. By training QML models on data representing different phases (e.g., order parameters, correlation functions, or even raw measurement outcomes from quantum simulations), it is possible to identify phase boundaries and characterize distinct quantum states~\cite{Carrasquilla2017machine,Van2017learning,Broecker2017machine,Carleo2019machine}.
For example, Quantum Support Vector Machines (QSVM) and
Variational Quantum Classifiers (VQC)
have been used with SHAP-driven feature selection to classify phases in
the Axial Next-Nearest Neighbor Ising (ANNNI) model. Another approach
combines classical shadows (an efficient quantum measurement technique) with
unsupervised machine learning (like K-Means clustering) to identify phase
transitions in models such as the ANNNI and Kitaev-Heisenberg models~\cite{Balthazar2023unsupervised}.
QML is also being applied to \textit{accelerate the discovery of new materials}
by learning complex relationships between material composition, structure,
and properties from large datasets, potentially guiding experimental
synthesis~\cite{Aspuru2005simulated,Montanaro2016quantum,Hutorchi2023overview,Romero2022quantum}.  Furthermore, QML techniques
can \textit{enhance quantum simulations} themselves, for instance,
by using neural networks to improve the accuracy or efficiency of
variational algorithms, as seen in the Variational Quantum-Neural
Hybrid Eigensolver (VQNHE)~\cite{Rsah2024variational} or Neural Quantum States (NQS)~\cite{Carleo2019machine,Lange2025transformerneural}.

The strengths of QML lie in its potential to process complex, high-dimensional quantum data and identify subtle patterns that might be missed by traditional analysis methods. Quantum computers could offer advantages in constructing powerful feature maps or kernels for certain types of data. However, the field faces significant challenges. \textit{Data encoding}, i.e., efficiently representing classical data in a quantum state or quantum data for classical processing, is a major hurdle. Training parameterized quantum circuits, a common component in many QML models, can suffer from \textit{barren plateaus}, similar to VQE. Demonstrating a clear \textit{quantum advantage} over highly optimized classical machine learning algorithms for practical problems remains an open question, and the \textit{qubit requirements} for handling large, real-world datasets can be substantial~\cite{Mcclean2018barren,Morais2025distinguishing}.

A significant trend emerging in QML for condensed matter physics is the pragmatic use of hybrid quantum-classical approaches. Rather than aiming for purely quantum machine learning solutions, many promising strategies involve a synergistic combination of quantum and classical resources. For example, quantum hardware can be used to generate data that is inherently difficult for classical computers to produce (e.g., expectation values from complex quantum states via classical shadows), which is then fed into powerful classical machine learning algorithms for analysis and model building. Conversely, classical machine learning techniques can assist quantum algorithms, such as using SHAP (Shapley Additive Explanations) for feature selection to reduce the input dimensionality for a quantum classifier, or employing neural networks to augment the expressive power or optimize the training of variational quantum algorithms like VQNHE. This hybrid paradigm acknowledges the current limitations of both purely classical and purely quantum approaches, seeking to leverage the respective strengths of each to achieve near-term progress in tackling complex condensed matter problems~\cite{Lundberg2017unified,Franco2025quantum}.

\begin{figure*}[t]
\centering
\includegraphics[width=0.86\linewidth]{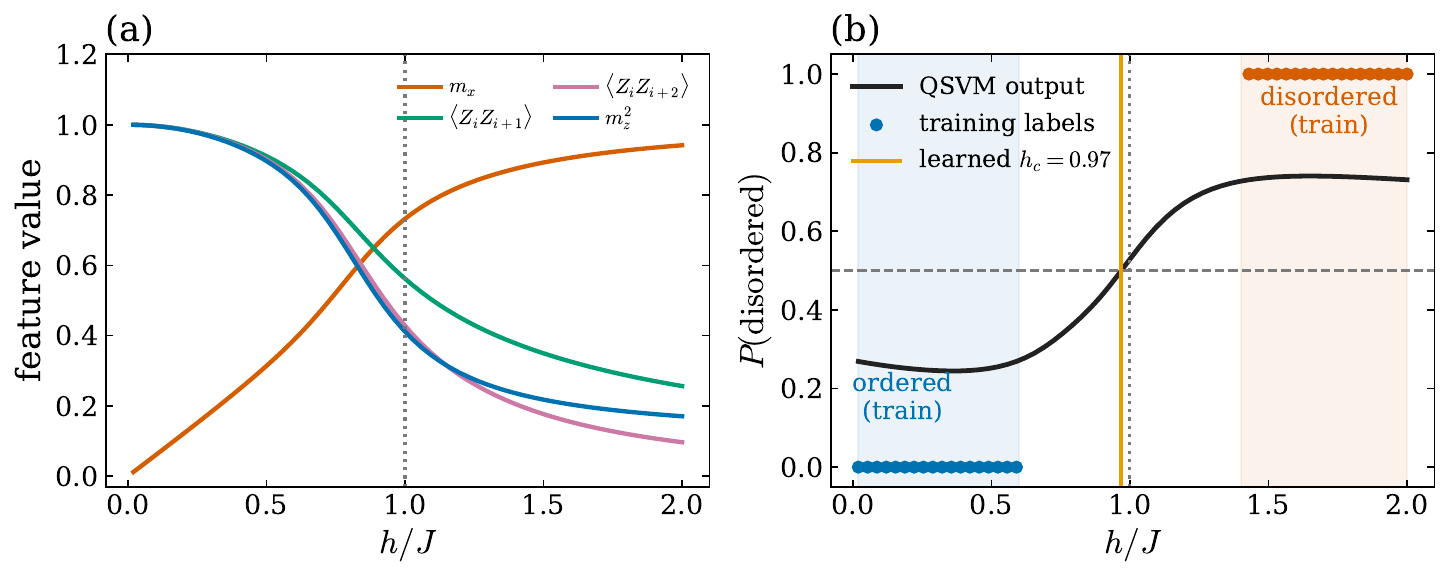}
\caption{Quantum-kernel classification of TFIM phases. (a)~Input features
extracted from $L=10$ ground states versus $h/J$. (b)~QSVM output probability
of the disordered phase, where the model is trained only on the shaded regions
($h/J\le0.6$, $h/J\ge1.4$) yet places the boundary (solid line) at
$h_c=0.97$, close to the exact value $h_c=J$ (dotted).}
\label{fig:qml}
\end{figure*}
\begin{figure*}[t]
\centering
\includegraphics[width=\linewidth]{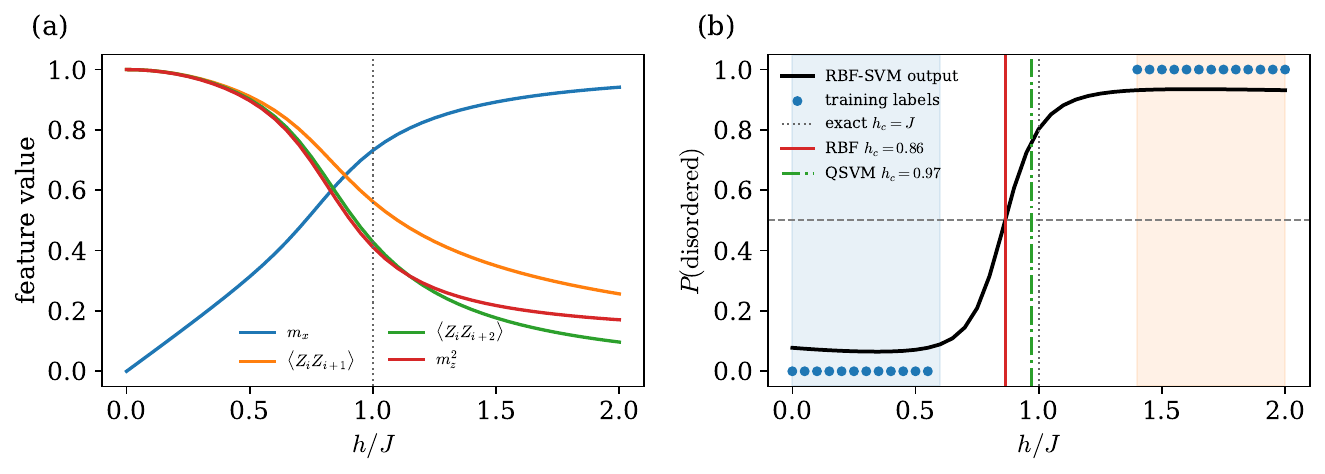}
\caption{Classical baseline for the phase-classification task of
Fig.~\ref{fig:qml}. (a)~The same four features extracted from $L=10$ TFIM
ground states versus $h/J$. (b)~Output of an ordinary radial-basis-function
(RBF) support-vector machine trained only on the shaded deep-phase regions
($h/J\le0.6$, $h/J\ge1.4$): the learned boundary (red) lands at
$h_c\approx0.86$, in the same range as the quantum kernel ($h_c=0.97$,
dash-dotted) and within finite-size rounding of the exact value $h_c=J$
(dotted). A classical kernel acting on identical observables is thus
competitive, and no decisive quantum advantage is seen on this task; the exact
boundary depends mildly on the SVM hyperparameters.}
\label{fig:qml_rbf}
\end{figure*}

To make the phase-classification application concrete, in the same spirit as
the ANNNI-model studies noted above, we implement a quantum-kernel support
-vector machine (QSVM). Ground states of the $L=10$ TFIM are obtained by sparse
exact diagonalization across the transition, and four physically measurable
features, the transverse magnetization, two longitudinal correlators, and the
structure factor,
\begin{equation}
\bm{x}=\big(m_x,\ \braket{Z_iZ_{i+1}},\ \braket{Z_iZ_{i+2}},\ m_z^2\big),
\label{eq:qmlfeatures}
\end{equation}
with $m_x$ and $m_z^2$ as in Eq.~\eqref{eq:orderparams}, are embedded in a
four-qubit feature map $\ket{\phi(\bm x)}=U_\phi(\bm x)\ket{0}^{\otimes4}$,
defining the quantum kernel and the kernelized decision function
\begin{equation}
K(\bm x,\bm x')=\big|\!\braket{\phi(\bm x')|\phi(\bm x)}\!\big|^2,\qquad
f(\bm x)=\mathrm{sign}\!\Big(\textstyle\sum_i\alpha_i y_i K(\bm x_i,\bm x)+b\Big)
\label{eq:qmlkernel}
\end{equation}
[Fig.~\ref{fig:qml}(a)]. Training the classifier \emph{only} on states deep in
each phase ($h/J\le0.6$ and $h/J\ge1.4$) and asking it to label the unseen
critical window, the learned decision boundary lands at $h_c=0.97$
[Fig.~\ref{fig:qml}(b)], within a few percent of the exact quantum critical
point $h_c=J$ despite finite-size rounding. The example demonstrates how a
quantum kernel acting on modest, experimentally accessible observables suffices
to detect a quantum phase transition. As a control, an ordinary
radial-basis-function SVM trained on the identical features and protocol locates
the transition at $h_c\approx0.86$ [Fig.~\ref{fig:qml_rbf}], comparable to the
quantum kernel and within finite-size rounding of $h_c=J$; on this task the
quantum kernel therefore offers no decisive advantage, in line with the cautious
view of quantum-machine-learning separations noted above.

\subsection{Quantum Simulation of Specific Models and Phenomena}

Beyond general algorithmic frameworks, a significant portion of quantum algorithm development in condensed matter physics focuses on simulating specific, highly relevant models and phenomena. These targeted simulations aim to provide insights into systems that are intractable for classical computers.

Several canonical models recur as the principal targets of these efforts. The Fermi-Hubbard model is central to understanding strongly correlated electron systems and is believed to capture the essential physics of high-temperature superconductivity~\cite{Stanisic2022observing,Arovas2022hubbard,Qin2022hubbard,Anderson1987resonating}, and quantum algorithms are being developed to find its ground states and excited states and to simulate its dynamics~\cite{Cade2020strategies,Stanisic2022observing,Google2020,Shao2024antiferromagnetic}, while software libraries such as \texttt{Qiskit Nature}~\cite{qiskit_nature_2023,qiskit_nature_paper} and \texttt{OpenFermion-Cirq}~\cite{Mcclean2020openfermion,Openfermion_tutorials} provide tools specifically for constructing and simulating Hubbard Hamiltonians. Spin models are equally fundamental for studying quantum magnetism, quantum phase transitions~\cite{Sachdev2011quantum}, and spin liquids~\cite{Savary2016quantum}, and they have deep connections to quantum information processing~\cite{nielsen2010quantum}. Within this family the \textit{Ising model}, in various dimensions and with transverse fields, is a workhorse for studying phase transitions and is often used as a benchmark for quantum annealers and QAOA~\cite{Sachdev2011quantum,Kadowaki1998quantum,Farhi2001quantum,Hauke2020,Shaydulin2023evidence}, the \textit{Heisenberg model} describes interacting quantum spins and is crucial for understanding magnetic materials and phenomena such as quantum entanglement~\cite{Mattis2006theory,Amico2008entanglement}, and the \textit{Kitaev model}, particularly on honeycomb lattices, is known for its exactly solvable spin-liquid ground state and its potential for realizing topological quantum computation through Majorana fermions~\cite{Kitaev1995quantum,Savary2016quantum,Hermanns2018physics}.

Beyond these lattice and spin systems, the Sachdev-Ye-Kitaev model of randomly interacting Majorana fermions has attracted significant attention because of its solvability in certain limits, its exhibition of quantum chaos, and its holographic connections to black-hole physics in string theory, so that its simulation on quantum computers could provide insight into these deep theoretical connections~\cite{Sachdev1993gapless,Kitaev2015simple,Rosenhaus2019introduction,Sachdev2017holographic,Garcia2021digital}. Quantum algorithms are likewise being designed to prepare, manipulate, and characterize topological phases of matter, which are robust against local perturbations and host exotic excitations such as anyons, including non-Abelian anyons such as Majorana zero modes and parafermions~\cite{nayak2008non,Kitaev2003fault}, and this effort extends to simulating Chern insulators, the quantum spin Hall effect, and fractional quantum Hall states, with the realization of topological qubits remaining a major goal for fault-tolerant quantum computing~\cite{Azad2022quantum}. A related frontier is the study of dynamical quantum phase transitions, the non-analytic behaviors in the return probability or Loschmidt echo that follow a quantum quench and signal critical phenomena in non-equilibrium quantum dynamics~\cite{Heyl2013dynamical,Heyl2018dynamical,Mueller2023}, for which quantum computers offer a particularly natural platform.

The choice of a specific condensed matter model often guides the selection of the most appropriate quantum algorithm and, consequently, the software tools. For instance, D-Wave's quantum annealers are inherently suited for optimization problems that can be mapped to Ising Hamiltonians, making them a natural choice for certain spin glass problems~\cite{King2023}. In contrast, gate-based SDKs like \texttt{Qiskit}, \texttt{Cirq}, and \texttt{PennyLane} offer greater flexibility for implementing algorithms like VQE or QPE to study the ground states or dynamics of models like the Fermi-Hubbard or Heisenberg Hamiltonians. This model-specificity implies that the development of quantum algorithm software must not only focus on general-purpose algorithmic frameworks but also provide specialized modules, interfaces, and pre-packaged routines tailored to these canonical condensed matter Hamiltonians to facilitate their study by physicists.

To turn the fermionic discussion above into concrete numbers, and to quantify
the mapping overhead that dominates the cost of simulating electrons, we
construct the half-filled Fermi-Hubbard Hamiltonian
\begin{equation}
H_{\rm FH}=-t\!\!\sum_{\langle ij\rangle,\sigma}\!\!\big(c^{\dagger}_{i\sigma}c_{j\sigma}
+\mathrm{h.c.}\big)+U\sum_i n_{i\uparrow}n_{i\downarrow}-\mu\sum_{i\sigma}n_{i\sigma},
\label{eq:hubbard}
\end{equation}
on clusters from two to ten sites and map it to qubits with both the
Jordan-Wigner (JW) and Bravyi-Kitaev (BK) encodings using \texttt{OpenFermion}.
The JW transform replaces each fermion by a Pauli string that enforces
antisymmetry through a parity tail,
\begin{equation}
c_j=\Big(\prod_{k<j}Z_k\Big)\frac{X_j+iY_j}{2},
\label{eq:jw}
\end{equation}
keeping local terms local on a chain but generating long strings in two
dimensions, whereas BK stores parity on logarithmic-weight sets of qubits and
trades string length for a denser update structure. The cost we quote is that of
a single first-order Trotter step,
\begin{equation}
e^{-iH\,dt}\approx\prod_a e^{-i c_a P_a\,dt},\qquad H=\sum_a c_a P_a,
\label{eq:trotter}
\end{equation}
in which a Pauli string of weight $w$ is synthesized by a CNOT staircase using
$2(w-1)$ two-qubit gates.
Table~\ref{tab:hubbard} reports, for each cluster, the exact ground-state energy
(our classical reference, from sparse diagonalization) together with the qubit
count, the number of Pauli terms, the maximum Pauli weight, and the two-qubit
-gate cost of a single first-order Trotter step. The comparison is more nuanced
than the common statement that one encoding is uniformly cheaper. On chains JW
keeps every mapped operator local ($w_{\max}=3$ independent of length) and is the
more economical choice, whereas in two dimensions the JW strings lengthen and BK
reduces both the operator weight and the gate count, for example $90$ versus
$104$ CNOTs for the $2\times2$ cluster [Fig.~\ref{fig:resources}(a,b)]. This is
exactly the Hamiltonian- and geometry-dependent trade-off that fermionic
simulation software must expose, and it sets concrete resource targets for the
algorithms of Sec.~\ref{sec:algorithms}.

\begin{table*}[t]
\caption{\label{tab:hubbard}Resource comparison of the Jordan-Wigner (JW) and Bravyi-Kitaev (BK) fermion-to-qubit mappings for the half-filled Fermi-Hubbard model ($t=1$, $U=4$, $\mu=U/2$, open boundaries). For each cluster we list the qubit count $N$, the exact ground-state energy $E_0$ (sparse diagonalization, shown where $N\le14$), the number of Pauli terms (identical for both mappings), and, for each mapping, the maximum Pauli weight $w_{\max}$ and the number of CNOTs in one first-order Trotter step $e^{-iH\,dt}$ (CNOT-staircase synthesis, all-to-all connectivity). Jordan-Wigner preserves locality on chains ($w_{\max}=3$) but produces long strings in two dimensions, where Bravyi-Kitaev reduces both the weight and the gate count.}
\begin{ruledtabular}
\begin{tabular}{lccccccc}
 & & & & \multicolumn{2}{c}{Jordan-Wigner} & \multicolumn{2}{c}{Bravyi-Kitaev}\\
\cline{5-6}\cline{7-8}
cluster & $N$ & $E_0/t$ & terms & $w_{\max}$ & CNOTs & $w_{\max}$ & CNOTs\\
$1\times2$ & 4 & $-4.828$ & 6 & 3 & 20 & 3 & 16\\
$1\times3$ & 6 & $-7.236$ & 11 & 3 & 38 & 5 & 42\\
$2\times2$ & 8 & $-10.103$ & 20 & 5 & 104 & 5 & 90\\
$1\times4$ & 8 & $-9.953$ & 16 & 3 & 56 & 5 & 62\\
$2\times3$ & 12 & $-15.619$ & 34 & 5 & 188 & 7 & 188\\
$1\times6$ & 12 & $-15.093$ & 26 & 3 & 92 & 7 & 118\\
$2\times4$ & 16 & n/a & 48 & 5 & 272 & 7 & 268\\
$3\times3$ & 18 & n/a & 57 & 7 & 402 & 7 & 388\\
$2\times5$ & 20 & n/a & 62 & 5 & 356 & 8 & 380\\
\end{tabular}
\end{ruledtabular}
\end{table*}

\subsection{Tensor Network Methods (Classical and Quantum-Inspired)}
Tensor Network (TN) methods~\cite{Orus2014practical} are a class
of powerful classical numerical techniques for simulating quantum many-body systems,
particularly those in low dimensions or exhibiting limited entanglement.
Prominent examples include Matrix Product States (MPS)
for $1$D systems and Projected Entangled Pair States (PEPS) for $2$D systems~\cite{Schollwock2011density,Eisert2010colloquium}. These methods
work by efficiently representing the quantum state or operators using a network of
interconnected tensors, thereby mitigating the exponential scaling of the full Hilbert
space for certain classes of states.

In the context of quantum algorithms for condensed matter physics, TNs play a
multifaceted role. Firstly, they serve as crucial \textit{benchmarks} for quantum
algorithms. By providing highly accurate classical solutions for specific models
(e.g., $1$D Hubbard or Heisenberg models), TN simulations can help validate the results
obtained from quantum hardware or quantum algorithm simulators, especially in the
NISQ era where quantum results are noisy and approximate~\cite{Bharti2022noisy,Yoshioka2024hunting}.
Secondly, TNs can \textit{inspire new quantum algorithms}. The structure and principles
of TNs, such as their efficient representation of entanglement, can inform the design
of novel quantum circuits and ansatze. A prime example is the Multiscale Entanglement
Renormalization Ansatz (MERA), a type of TN that has been adapted for implementation
on quantum computers to study quantum critical phenomena and phase transitions~\cite{Vidal2008class,Kim2017robust,Evenbly2019representation,Orus2014practical}.

Furthermore, TNs are increasingly being integrated into the quantum algorithm
development pipeline. They can be used for \textit{classical simulation of
quantum computations}, particularly for circuits with specific structures or
limited entanglement, sometimes simulating experiments that were previously
believed to be beyond classical reach. TN methods can also be employed
to \textit{initialize quantum computations} by preparing an approximate ground
state classically, which is then loaded onto the quantum computer as a starting
point for algorithms like VQE or QPE, potentially speeding up convergence.
The application of TNs extends to quantum circuit synthesis and even quantum error correction~\cite{Pan2022solving,Berezutskii2025tensor}.

This evolving role indicates a synergistic relationship between TN methods and quantum computing, rather than a purely competitive one. While TNs have their own limitations (e.g., difficulty with highly entangled states in 2D and 3D, or simulating long-time dynamics), their strengths in representing certain important classes of quantum states make them invaluable. Future quantum software ecosystems are likely to feature more sophisticated integration of TN tools, not only for classical pre-processing, benchmarking, and post-processing, but also as components within hybrid quantum-classical algorithms or as direct inspiration for quantum circuit design tailored to condensed matter problems~\cite{Daley2022,Bauer2020quantum,Tubman2020warm}.

\begin{figure*}[t]
\centering
\includegraphics[width=\linewidth]{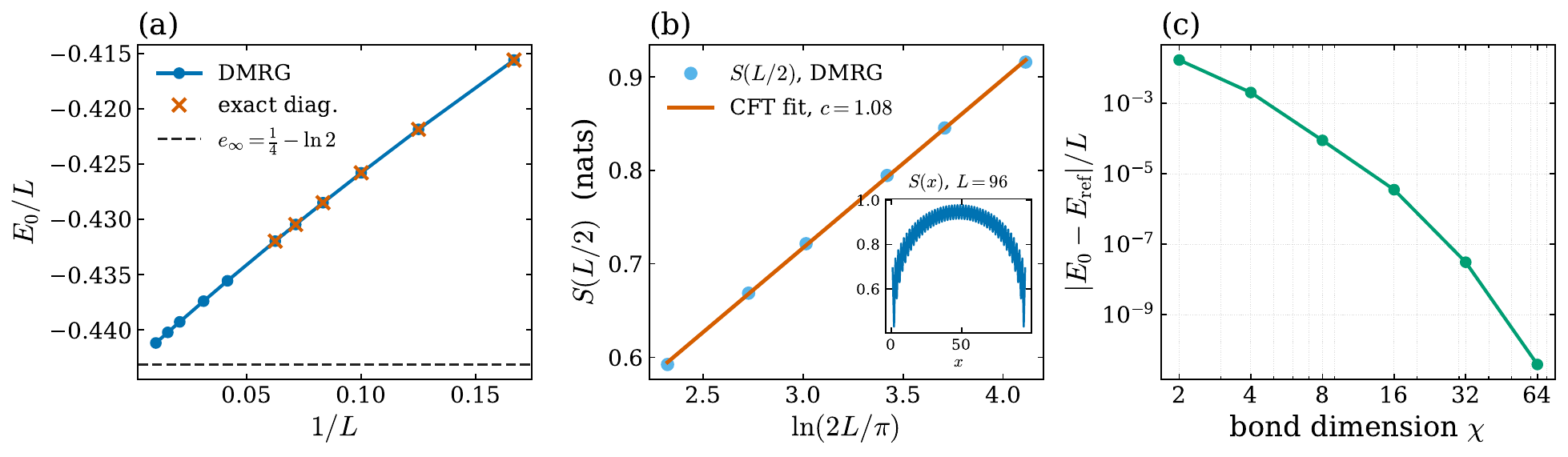}
\caption{DMRG study of the spin-$\tfrac12$ Heisenberg chain. (a)~Ground-state
energy per site versus $1/L$ from DMRG and exact diagonalization, approaching
$e_\infty=\tfrac14-\ln2$ (dashed). (b)~Finite-size scaling of the midpoint
entanglement entropy $S(L/2)$ against $\ln(2L/\pi)$, whose slope gives the
central charge $c\simeq1$, with the inset showing the full entanglement profile $S(x)$ for $L=96$.
(c)~Ground-state energy error versus MPS bond dimension $\chi$, showing
controlled, systematically improvable convergence.}
\label{fig:tn}
\end{figure*}

As a representative tensor-network calculation, and the classical benchmark
invoked throughout this review, we use DMRG with matrix-product states (via
\texttt{quimb}) for the spin-$\tfrac12$ antiferromagnetic Heisenberg chain,
$H=\sum_i \bm{S}_i\!\cdot\!\bm{S}_{i+1}$. The method represents the state as a
chain of rank-three tensors of bond dimension $\chi$,
\begin{equation}
\ket{\psi}=\sum_{\{s\}}A^{s_1}A^{s_2}\cdots A^{s_L}\ket{s_1 s_2\cdots s_L},
\label{eq:mps}
\end{equation}
which caps the representable entanglement at $S\le\ln\chi$ and is therefore exact
for area-law states but only systematically approximate at a logarithmic
critical point. The ground-state energy per site
agrees with exact diagonalization to machine precision for $L\le16$ and
converges toward the Bethe-ansatz value $e_\infty=\tfrac14-\ln2$ as the chain
grows [Fig.~\ref{fig:tn}(a)]. The half-chain entanglement entropy obeys the
Calabrese-Cardy logarithmic law for an open chain,
\begin{equation}
S(\ell)=\frac{c}{6}\,\ln\!\Big[\frac{2L}{\pi}\sin\frac{\pi\ell}{L}\Big]+s_1,
\label{eq:cc}
\end{equation}
and finite-size scaling of the midpoint entropy $S(L/2)$
yields a central charge $c\simeq1$ [Fig.~\ref{fig:tn}(b)], correctly
identifying the gapless Tomonaga-Luttinger-liquid universality class, with the
characteristic entanglement ``arch'' shown in the inset. Finally, the
ground-state energy error falls by nine orders of magnitude as the bond
dimension increases from $\chi=2$ to $\chi=64$ [Fig.~\ref{fig:tn}(c)],
exemplifying the systematically improvable accuracy that makes tensor networks
the gold-standard classical reference against which the quantum algorithms
above are measured.

\section{Classical Algorithmic Approaches and Their Limitations}

Before the advent of quantum computing, and still predominantly today, the study of condensed matter systems has relied on a variety of classical computational methods~\cite{Fehske2007computational,Martin2020electronic}. These techniques have achieved considerable success in explaining and predicting material properties. Prominent classical approaches include several complementary methods. \textit{Exact Diagonalization} (ED)~\cite{Sandvik2010computational} directly diagonalizes the Hamiltonian matrix for small systems. \textit{Quantum Monte Carlo} (QMC) methods use stochastic sampling to estimate quantum mechanical expectation values. \textit{Density Functional Theory} (DFT)~\cite{Jones2015density} is a widely used method for electronic structure calculations based on the electron density. \textit{Dynamical Mean-Field Theory} (DMFT)~\cite{Georges1996dynamical} maps a lattice problem onto a self-consistent quantum impurity problem and is particularly effective for strongly correlated systems. Finally, \textit{Tensor Network} (TN) methods provide efficient representations for certain classes of quantum states, especially in one dimension.

These classical methods have enabled significant breakthroughs. For instance, DFT has become a workhorse for materials design~\cite{Jones2015density,Hasnip2014density}, while DMFT has provided crucial insights into phenomena like the Mott metal-insulator transition~\cite{Georges1996dynamical}. TN methods excel at describing ground states of gapped $1$D systems and have been instrumental in understanding entanglement in quantum matter~\cite{Eisert2010colloquium}.

However, each of these classical approaches faces inherent limitations when applied to the full spectrum of challenging condensed matter problems. ED is severely restricted by the exponential growth of the Hilbert space, typically limited to a few tens of particles~\cite{Fehske2007computational}. QMC methods, while powerful for bosonic systems or specific fermionic problems, often suffer from the "fermion sign problem" (or "minus sign problem") when applied to generic interacting fermionic systems at low temperatures or with frustration, leading to an exponential increase in computational cost to achieve a desired accuracy~\cite{Troyer2005computational}. DFT, in its common approximations (like LDA or GGA), struggles with strongly correlated materials where electron-electron interactions are dominant and cannot be treated as a simple perturbation~\cite{Anisimov1991band}. While DMFT addresses strong correlations, its accuracy for finite-dimensional systems depends on the approximation of the impurity solver and the treatment of non-local correlations~\cite{Georges1996dynamical}. Classical TN methods, while very successful for 1D systems and certain 2D systems with limited entanglement, face an exponential increase in computational cost when dealing with highly entangled states in higher dimensions or long-time dynamics~\cite{Orus2014practical}.

The intractability of the Schrödinger equation in its original form for most many-body systems and these specific limitations of classical algorithms directly motivate the development of quantum algorithms. For example, the fermion sign problem in QMC is a primary driver for exploring quantum simulations of fermionic Hamiltonians, as quantum computers naturally handle fermionic statistics (after appropriate mapping to qubits). The challenge of representing highly entangled states in classical TNs for $2$D and $3$D systems is another area where quantum computers, which can inherently sustain high levels of entanglement, are expected to offer an advantage.

The pursuit of "quantum advantage" involves identifying problems where quantum algorithms can provably and practically outperform the \textit{best} known classical algorithms. This often targets a super-quadratic speedup. However, rigorously establishing such an advantage is complicated by the continuous improvement of classical algorithms and the difficulty in proving classical hardness assumptions, such as those related to the P versus NP problem. The limitations of classical methods thus define the landscape of opportunities for quantum computation in condensed matter physics, guiding the selection of target problems and the design of quantum algorithms intended to overcome these specific classical bottlenecks~\cite{Troyer2005computational}.

\section{Software Development Kits (SDKs) and Libraries for Condensed Matter Physics}

The translation of theoretical quantum algorithms into executable programs for quantum hardware or simulators is facilitated by a growing ecosystem of Software Development Kits (SDKs) and specialized libraries. These tools are indispensable for researchers and developers in condensed matter physics aiming to explore quantum computational approaches.

\subsection{Overview of the Quantum Software Ecosystem}

Quantum SDKs provide the necessary abstractions and functionalities to design,
optimize, and run quantum programs. Key objectives of these software tools include
the implementation of diverse quantum algorithms, the incorporation of error mitigation
and, eventually, error correction schemes, the ability to simulate quantum circuits on
classical computers (often with noise models), and the capability to execute these
circuits on actual quantum hardware~\cite{King2023}. A typical quantum software stack includes
high-level quantum programming languages or Python-based interfaces for circuit
construction, transpilers for optimizing circuits and mapping them to specific
hardware constraints, sophisticated classical simulators, and backend interfaces
to various quantum processing units (QPUs). The choice of SDK often depends on
factors such as the target quantum hardware, the specific class of problems being
addressed (e.g., optimization, ground-state determination, dynamics), the desired
level of control over the quantum operations, and the user's familiarity with the
programming paradigm. As the field matures, interoperability between different SDKs
and tools, for example through common intermediate representations or libraries
like \texttt{OpenFermion}, is becoming increasingly important.

A critical underlying trend in software development for condensed matter
applications is the focus on "fermionic simulation libraries beyond
mappers"~\cite{Yu2025clifford}. Standard fermion-to-qubit mappings
like Jordan-Wigner (JWT), Bravyi-Kitaev (BKT)~\cite{Seeley2012bravyi}, or Parity transformations
are essential for representing fermionic Hamiltonians on qubit-based
quantum computers. However, these mappings can introduce significant
overhead in terms of the number of qubits required or the complexity
(e.g., Pauli weight, non-locality) of the resulting qubit operators,
which translates to deeper and noisier circuits~\cite{Jiang2020optimal}. Consequently,
there is substantial research into developing more efficient mapping
techniques tailored to specific Hamiltonians or hardware architectures.
This includes optimizing existing mappings by reordering fermionic modes,
using ancilla qubits to reduce operator complexity, or even exploring
analog fermionic quantum simulators that might circumvent traditional qubit
mappings altogether. The Fermionic Quantum Emulator (FQE) is an example
of a library designed for efficient simulation of fermionic dynamics, taking
advantage of symmetries~\cite{Rubin2021fermionic}. These efforts are crucial for making quantum
simulations of fermionic condensed matter systems practical on near-term and future quantum devices.

\subsection{Qiskit (IBM)}
\texttt{Qiskit}, developed by IBM, is a comprehensive open-source Python-based
SDK for quantum computing~\cite{qiskit_nature_2023,qiskit_nature_paper}. It
provides a rich set of tools for creating quantum circuits, transpiling them
for optimal execution on different backends, and running them on high-performance
classical simulators or IBM's cloud-accessible quantum hardware.  \texttt{Qiskit Aer}
is its primary simulation engine, offering various simulation methods (statevector,
density matrix, unitary), support for realistic noise modeling, and built-in error
mitigation techniques. A notable feature is the \texttt{Qiskit Runtime}
environment, which allows for more efficient execution of iterative algorithms by
co-locating classical computation closer to the quantum hardware, thereby reducing
latency \texttt{Qiskit} also includes tools like \texttt{Benchpress} for
benchmarking quantum software performance and boasts a modular architecture
designed for extensibility. Recent developments include AI-enhanced transpilation services.

For condensed matter physics, \textit{Qiskit Nature} is a key application module.
It supports solving problems in natural sciences, including the construction and
simulation of lattice models like the Ising and Fermi-Hubbard models.
The \textit{Qiskit Algorithms} library provides implementations of fundamental
quantum algorithms such as VQE (including variants like \texttt{AdaptVQE}), QPE,
and time evolution algorithms. \texttt{Qiskit} has been used, for example,
to demonstrate the simulation of Ising model time evolution on IBM hardware,
highlighting its capabilities for dynamic simulations~\cite{Qiskit_algorithms2023,Qiskit_textbook}.

\texttt{Qiskit} offers extensive tutorials, including those
for defining various lattice structures (\texttt{LineLattice},
\texttt{SquareLattice}) and working with the Fermi-Hubbard model.
There are also examples illustrating the implementation of QPE for eigenvalue problems~\cite{Qiskit_algorithms2023,Qiskit_textbook}.

\subsection{Cirq (Google)}

\texttt{Cirq}~\cite{cirq_developers_2024} is another open-source Python library, developed by Google,
primarily focused on creating, manipulating, and optimizing quantum circuits
for Noisy Intermediate-Scale Quantum (NISQ) devices. It provides
useful abstractions for dealing with the specific details and constraints
of quantum hardware. A key feature is the Quantum Virtual Engine (QVM),
which allows users to simulate circuits with an interface that mimics real hardware,
including realistic noise models. \texttt{Cirq} includes built-in simulators for wave
functions and density matrices, and it integrates with the high-performance \texttt{qsim} simulator~\cite{cirq_developers_2024}.

\texttt{Cirq} has been extensively used in Google's quantum hardware
experiments relevant to condensed matter physics. These include simulations
of the Fermi-Hubbard model, studies of Kardar-Parisi-Zhang (KPZ) universality
in 1D Heisenberg spin chains, and measurements of out-of-time-ordered correlators
(OTOCs) to probe quantum scrambling. The \textit{OpenFermion} library, which provides
tools for simulating fermionic systems, is commonly used in conjunction with \texttt{Cirq}
(\texttt{OpenFermion-Cirq}) for problems in quantum chemistry and condensed matter physics,
including Hubbard model simulations. The \textit{Fermionic Quantum Emulator (FQE)} also
integrates with \texttt{Cirq}, offering efficient methods for simulating fermionic
dynamics by exploiting symmetries~\cite{cirq_developers_2024}.

Tutorials and examples for \texttt{Cirq} include detailed walkthroughs of
Fermi-Hubbard experiments, demonstrations of Hubbard model simulations using
\texttt{OpenFermion-Cirq}, and guidance on implementing algorithms like QPE~\cite{cirq_developers_2024}.

\subsection{PennyLane (Xanadu)}

\texttt{PennyLane}, developed by Xanadu~\cite{Pennylane_documentation}, is a Python-based,
cross-platform software framework designed for differentiable
programming of quantum computers. Its core strength lies in
facilitating quantum machine learning (QML), quantum chemistry,
and the optimization of hybrid quantum-classical computations.
A key feature of \texttt{PennyLane} is its ability to compute
gradients of variational quantum circuits in a way that is
compatible with classical automatic differentiation libraries
such as PyTorch, TensorFlow, and JAX. It employs a flexible
plugin system, allowing it to interface with a wide array of
quantum hardware backends (including IBM Quantum, Amazon Braket,
Google Quantum AI) and classical simulators~\cite{Pennylane_documentation,Pennylane_qchem_module,Bergholm2022pennylane}.

For condensed matter applications, \texttt{PennyLane}
offers tools for quantum chemistry that are applicable
to material systems, such as building molecular Hamiltonians
and preparing initial states. The \texttt{qml.spin}
module includes functions like \texttt{heisenberg} for
constructing Hamiltonians for the Heisenberg model
on various lattice geometries. \texttt{PennyLane} also
provides datasets, such as one for the Bose-Hubbard model,
which can be used for developing and benchmarking QML algorithms.
It supports various fermionic mapping schemes, including Jordan-Wigner, Parity, and Bravyi-Kitaev transformations, along with techniques for qubit tapering~\cite{Pennylane_documentation,Pennylane_qchem_module,Bergholm2022pennylane}.

\texttt{PennyLane}'s documentation and demo library feature
numerous examples relevant to condensed matter physics. These
include a challenge on simulating the noisy Heisenberg model,
tutorials on Quantum Phase Estimation, and examples of fermionic
simulations, including how to map fermionic operators
to qubits and prepare Hartree-Fock states~\cite{Pennylane_documentation,Pennylane_qchem_module,Bergholm2022pennylane}.

\subsection{Q\# and Azure Quantum Development Kit (Microsoft)}

Q\# (Q-sharp) is a domain-specific programming language developed
by Microsoft for writing and running quantum algorithms~\cite{microsoft_qdk_preview_blog,microsoft_qsharp_overview}.
It is part of the Azure Quantum Development Kit (QDK). Q\# is designed
with scalability in mind, aiming to support future large-scale,
fault-tolerant quantum applications. It offers a higher level of
abstraction compared to circuit-centric Python libraries, meaning
developers often work with operations and functions rather than
directly manipulating quantum states or circuits. Key features include
compiler-generated adjoint and controlled versions of operations,
and rich classical control flow constructs that can be seamlessly
integrated with quantum operations.  Microsoft has recently released
a preview of a new QDK, largely rewritten in Rust, which promises
significant improvements in installation size, speed, and usability,
including full browser support. Azure Quantum provides cloud-based
access to various quantum hardware and simulators~\cite{microsoft_qdk_preview_blog,microsoft_qsharp_overview}.

Microsoft's QDK includes a \textit{chemistry library} that is highly
relevant for materials science and condensed matter physics. This
library provides state-of-the-art Q\# implementations for Hamiltonian
simulation methods, including Trotterization and Qubitization techniques,
as well as state preparation methods. It supports the estimation
of ground and excited state energies and interfaces with computational
chemistry tools through the Broombridge schema, an open-source YAML-based
format for representing molecular Hamiltonians. While many examples
focus on molecular chemistry, the underlying Hamiltonian simulation
techniques are broadly applicable to condensed matter systems that can
be described by similar fermionic or spin Hamiltonians~\cite{microsoft_qdk_preview_blog,microsoft_qsharp_overview}.

Examples and tutorials provided with the QDK demonstrate Hamiltonian
evolution for molecules like water and showcase the use of the Broombridge schema.
While direct, advanced condensed matter tutorials for models like Hubbard or
Heisenberg are less explicitly detailed in the provided snippets compared to
\texttt{Qiskit} or \texttt{Cirq}, the foundational tools for Hamiltonian
simulation in Q\# are applicable. Recent research using Q\# or relevant
to its ecosystem explores initial state preparation for the Fermi-Hubbard
model by leveraging the Heisenberg model and the Quantum Imaginary
Time Evolution (QITE) algorithm for strongly correlated systems~\cite{Motta2020determining}.

\section{Challenges in Quantum Algorithm Software for Condensed Matter Physics}

Despite significant progress, the development and application of quantum
algorithm software for condensed matter physics face numerous interconnected
challenges. These span hardware limitations, algorithmic scalability, the
complexities of error handling, software design principles, and the overarching
goal of demonstrating practical quantum advantage.

\subsection{Hardware Limitations}

The capabilities of current quantum hardware remain a primary bottleneck~\cite{Preskill2018,Bharti2022noisy,Kandala2017}.
A first limitation concerns qubit quality and coherence. Qubits are highly susceptible to environmental noise, leading to decoherence, meaning the loss of quantum information, and to errors in gate operations, and the resulting limited coherence times restrict the depth of quantum circuits that can be reliably executed, which is a major constraint for many complex algorithms such as QPE. Achieving high-fidelity gate operations is therefore crucial, since errors accumulate rapidly in multi-gate circuits. A second limitation concerns qubit quantity and scalability. While the number of qubits in processors is increasing, current systems are firmly in the NISQ (Noisy Intermediate-Scale Quantum) era, typically featuring tens to a few hundred noisy qubits, whereas simulating condensed matter systems of practical interest often requires significantly more qubits, potentially millions for fault-tolerant computations of complex materials, so that scaling up while maintaining high qubit quality and connectivity remains a formidable engineering challenge. A third limitation is connectivity. Most current quantum hardware platforms have limited qubit connectivity, meaning that not all pairs of qubits can directly interact, and this necessitates the use of SWAP gates to move quantum information across the chip, which adds to the circuit depth and introduces additional error sources, further complicating the simulation of many-body Hamiltonians that often involve non-local interactions.

\subsection{Algorithm Scalability and Efficiency for Condensed Matter Problems}

Many promising quantum algorithms, when analyzed for practical
condensed matter problems, reveal substantial resource requirements
in terms of qubit numbers and circuit depth. The "curse of dimensionality,"
though potentially tamed by quantum mechanics, can reappear in algorithmic
complexity if not carefully managed.
A particularly significant challenge for condensed matter physics is
the simulation of \textit{fermionic systems}. Mapping fermionic creation
and annihilation operators to qubit operators (e.g., via Jordan-Wigner,
Bravyi-Kitaev, or Parity transformations) often results in qubit Hamiltonians
with long, non-local Pauli strings. Implementing these non-local terms on
hardware with limited connectivity requires many SWAP gates, increasing
circuit depth and error rates. Developing more efficient fermion-to-qubit
mappings or alternative simulation strategies is crucial~\cite{Bauer2020quantum}.

\begin{figure*}[t]
\centering
\includegraphics[width=\linewidth]{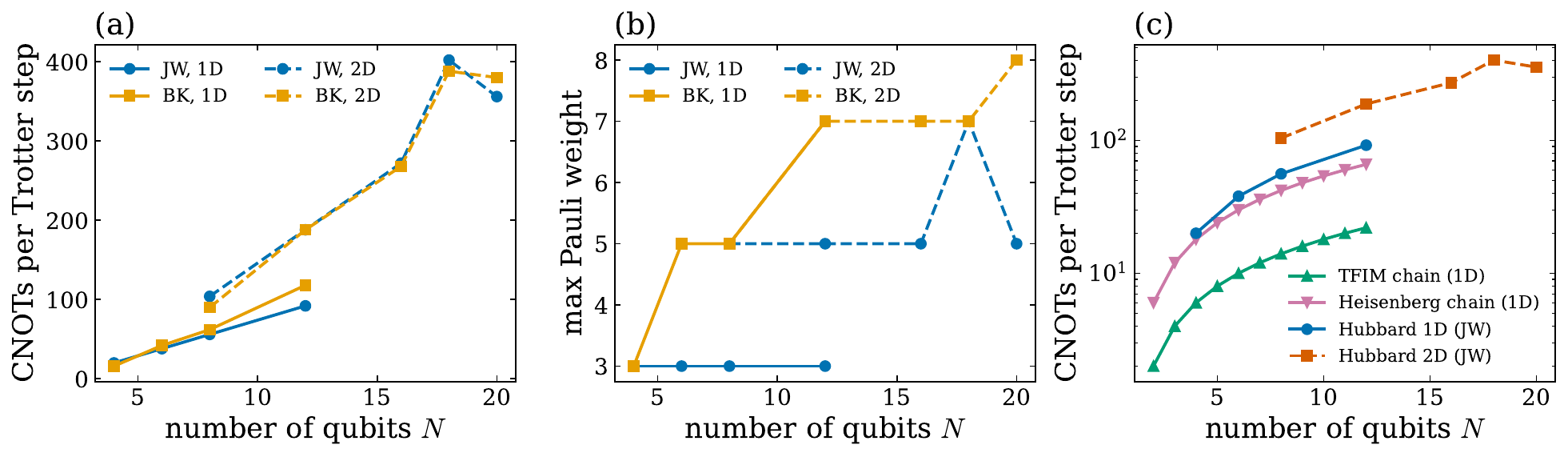}
\caption{Resource scaling of Trotterized simulation. (a)~CNOTs per first-order
Trotter step for the Fermi-Hubbard model under the Jordan-Wigner (JW) and
Bravyi-Kitaev (BK) mappings, separated into one- and two-dimensional clusters.
(b)~Maximum Pauli weight (operator locality) versus qubit number, with JW staying local
($w_{\max}=3$) on chains but not in two dimensions. (c)~CNOTs per Trotter step
across models, contrasting the linear growth of one-dimensional spin chains and
Hubbard clusters with the steeper cost of the two-dimensional Fermi-Hubbard
model. Counts assume the standard CNOT-staircase synthesis and all-to-all
connectivity, and are therefore lower bounds on hardware requiring SWAP routing.}
\label{fig:resources}
\end{figure*}

To make these resource statements quantitative rather than qualitative,
Fig.~\ref{fig:resources} tracks the cost of a single Trotter step across models
and mappings. The two-qubit-gate count of the canonical spin chains grows only
linearly in the number of qubits [$2(L-1)$ for the transverse-field Ising chain
and $6(L-1)$ for the Heisenberg chain], and one-dimensional Hubbard clusters
remain comparably cheap. The genuinely demanding case is the two-dimensional
Fermi-Hubbard model, whose non-local mapped operators drive a markedly steeper
growth [Fig.~\ref{fig:resources}(c)]. Such model- and geometry-resolved resource
curves, rather than generic qubit-count estimates, are what is needed to decide
where near-term hardware can realistically operate and to direct compiler effort,
for instance SWAP reduction, to where it matters most.

\subsection{Error Mitigation and Fault Tolerance}

Given the noisy nature of NISQ devices, \textit{error mitigation}
techniques are essential for extracting meaningful results. These methods,
which include techniques like zero-noise extrapolation, probabilistic
error cancellation, and readout error correction, aim to reduce the
impact of noise at the software level, often by running additional
quantum circuits and performing classical post-processing. While
beneficial, error mitigation typically incurs a significant overhead
in terms of both quantum runtime and classical computation~\cite{cai2023quantum,Endo2021hybrid}.

\begin{figure*}[t]
\centering
\includegraphics[width=\linewidth]{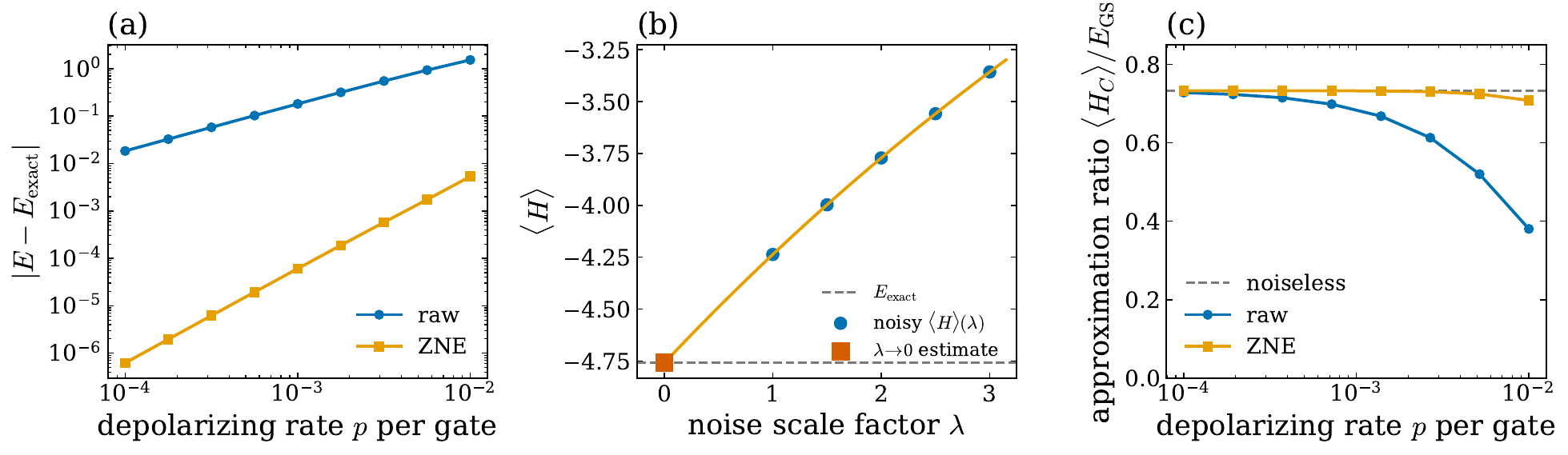}
\caption{Noise and error mitigation for the worked examples under a
gate-dependent depolarizing model (rate $p$ per single-qubit gate and $2p$ per
wire per two-qubit gate). (a)~VQE ground-state energy error versus $p$ for the
$L=4$ transverse-field Ising model, without mitigation and with exponential
zero-noise extrapolation (ZNE). (b)~ZNE at $p=3\times10^{-3}$, where the noisy energy
$\langle H\rangle(\lambda)$ at scaled noise $\lambda\in\{1,2,3\}$ is extrapolated
to $\lambda\!\to\!0$, recovering the exact value. (c)~QAOA approximation ratio
versus $p$ for the $L=6$ Ising spin glass, raw and ZNE-mitigated, against the
noiseless value (dashed).}
\label{fig:noise}
\end{figure*}

To show that these techniques are not merely aspirational for condensed-matter
circuits, we place the VQE and QAOA examples of Sec.~\ref{sec:algorithms} on a
density-matrix simulator with a gate-dependent depolarizing model and apply
zero-noise extrapolation (ZNE), with the result shown in Fig.~\ref{fig:noise}.
For the transverse-field Ising VQE the raw energy error grows with the per-gate
error rate $p$ and reaches order unity by $p\sim10^{-2}$, whereas exponential ZNE
recovers the ground-state energy to within about $10^{-3}$ over the entire
realistic range $10^{-4}\le p\le10^{-2}$, an improvement of two to four orders of
magnitude [Fig.~\ref{fig:noise}(a,b)]. The same mitigation rescues the QAOA
approximation ratio for the $L=6$ spin glass, which without mitigation collapses
from its ideal value $0.73$ to $0.38$ at $p=10^{-2}$ but is restored to $0.71$ by
ZNE [Fig.~\ref{fig:noise}(c)]. The demonstration also delineates the limits of
the method. Because ZNE is an extrapolation, its accuracy degrades once the
circuit approaches full decoherence, so error mitigation extends, but does not
replace, the eventual need for fault tolerance.

The long-term solution to hardware noise is \textit{fault-tolerant
quantum computing} (FTQC), which involves encoding logical qubits
using many physical qubits and actively correcting errors. However,
FTQC demands extremely low physical error rates (below a certain threshold)
and imposes a very large qubit overhead. The transition from current
NISQ approaches to FTQC is a major ongoing research effort.
Initiatives like Microsoft's development of topological qubits
(e.g., the Majorana 1 chip) aim to create qubits that are inherently
more robust against certain types of errors, potentially easing the path to fault tolerance~\cite{Terhal2015quantum}.

\subsection{Software Abstraction and User-Friendliness}

Developing effective quantum software involves a delicate balance.
On one hand, there is a need for high-level abstractions that allow
condensed matter physicists (who may not be quantum computing experts)
to easily express their problems and algorithms. On the other hand,
achieving optimal performance on NISQ devices often requires low-level
control over circuit compilation, pulse shaping, and error mitigation
strategies, which necessitates more detailed hardware knowledge.
Creating software stacks that cater to both types of users and
manage this complexity effectively is an ongoing challenge~\cite{laflamme_et_al_2024_benchmarking}.

\subsection{Demonstrating Quantum Advantage}

A central goal in the field is to demonstrate
\textit{quantum advantage}, where a quantum computer
solves a scientifically or industrially relevant problem
significantly faster or more accurately than the best
known classical algorithms on the most powerful supercomputers.
For condensed matter physics, this means tackling problems like
determining the ground state of a complex Hamiltonian or
simulating quantum dynamics in regimes inaccessible to
classical methods. However, rigorously proving quantum advantage
is difficult. Classical algorithms are continually improving, and
classical hardness assumptions (like P vs NP)
are hard to resolve definitively. Furthermore, the output of
a useful quantum algorithm should ideally be verifiable,
or at least repeatable by another quantum computer,
to build confidence in the results~\cite{Daley2022,Harrow2017quantum,Carolan2020verifying,Eisert2020certification}.

The journey towards practical quantum advantage in condensed matter physics
is not solely about constructing larger quantum computers. It critically
depends on the development of "smarter" software. This software must be
capable of maximally exploiting the limited and noisy quantum resources
available in the NISQ era. This involves sophisticated techniques for optimized
compilation (mapping abstract circuits to physical qubit layouts and native gate sets),
advanced error mitigation strategies deeply integrated into the workflow, and the
effective orchestration of hybrid quantum-classical computations where classical
resources assist the quantum processor. Software tools like Q-CTRL's Fire Opal~\cite{qctrl_fire_opal},
which focus on abstracting hardware complexity and automating error suppression,
exemplify this trend. The software layer is thus becoming a pivotal enabler for
extracting scientific value from current and near-future quantum devices~\cite{laflamme_et_al_2024_benchmarking,Bharti2022noisy,cai2023quantum}.

Another inherent tension in quantum software development is between the
desire for universal, hardware-agnostic algorithm descriptions and the pressing
need for hardware-specific optimizations to achieve the best possible performance
in the resource-constrained NISQ era. While portability and high-level programming
are attractive goals, the reality is that the performance of a quantum algorithm can
vary dramatically depending on how well it is tailored to the specific characteristics
of the underlying quantum processor, such as its qubit connectivity, native gate set,
and dominant error channels. For example, minimizing the number of SWAP gates during
circuit compilation is a crucial hardware-specific optimization.  This suggests that
mature quantum software stacks will likely need to employ layered abstractions~\cite{Bharti2022noisy,laflamme_et_al_2024_benchmarking}. High-level,
physics-oriented interfaces could allow condensed matter scientists to define their
problems naturally, while lower-level compilation and optimization passes,
increasingly automated and AI-driven, would handle the hardware-specific details.

\section{Future Trajectories and Outlook}

The field of quantum algorithm software for condensed matter physics
is rapidly evolving, driven by concurrent advances in quantum hardware,
algorithmic theory, and software engineering. Future trajectories point
towards more powerful algorithms, sophisticated software tools, and a
deeper integration of quantum computing into the scientific discovery
process for materials and quantum phenomena.

\subsection{Anticipated New Algorithms and Algorithmic Improvements}

One significant long-term goal is topological quantum computation, namely the realization of fault-tolerant quantum computation using topological qubits, which are intrinsically protected against certain types of errors, and research into realizing and manipulating non-Abelian anyons such as Majorana zero modes in condensed matter systems or parafermions is therefore crucial. Microsoft's development of the Majorana 1 chip based on topological superconductors is a notable step in this direction, and future algorithms will likely focus on efficient ways to perform braiding operations and to exploit the unique properties of these exotic particles for computation, with quantum simulation of parafermions in superconducting circuits forming another active area~\cite{nayak2008non,Kitaev2003fault}.

Improved variational and estimation algorithms are also anticipated. For VQE one expects continued development of more robust and efficient ansatze tailored to specific condensed matter Hamiltonians, better classical optimization strategies to navigate complex energy landscapes and avoid barren plateaus, and more effective integration of advanced error mitigation techniques, with the Knowledge Distillation Inspired VQE (KD-VQE) showing promise for improved convergence~\cite{Tilly2022variational}. For QPE the push toward making the algorithm practical on near-term devices will continue, with further refinements of control-free QPE methods that leverage classical phase retrieval and of generalized QPE (GQPE) for accessing nonlinear response properties~\cite{de_lima_de_farias_2024_qpe}. For quantum annealing, enhancements such as Learning-Driven Annealing (LDA) aim to improve the performance of quantum annealers by adaptively modifying problem Hamiltonians to mitigate hardware constraints~\cite{Grajciar2024learning}.

Quantum machine learning is expected to yield more powerful models for tasks such as quantum phase classification, automated materials discovery from large datasets, and the guiding or accelerating of quantum simulations, and scalable Quantum Architecture Search (QAS) techniques will be crucial for automatically designing efficient parameterized quantum circuits for QML and VQE~\cite{Franco2025quantum,Cerezo2022challenges}. Although much focus has been on ground-state properties, there is growing interest in algorithms for simulating quantum dynamics, transport phenomena, and non-equilibrium states such as dynamical quantum phase transitions~\cite{Miessen2023quantum}. Finally, research continues into identifying fundamental quantum subroutines beyond the Quantum Fourier Transform, and one avenue explores deriving new quantum primitives from the study of fast classical numerical transforms, such as Hermite, orthogonal-polynomial, and generalized Fourier transforms, which could lead to novel fast-forwarding algorithms for quantum simulation~\cite{bnl_novel_algorithms}.

\subsection{Expected Software Improvements}

Several improvements in software are anticipated. Enhanced abstraction layers will likely feature more sophisticated Quantum Abstract Machines (QAMs) and higher-level programming paradigms that allow users to express quantum algorithms more naturally while abstracting away many of the low-level hardware complexities related to qubit mapping, gate decomposition, and pulse control. Performance optimization is expected to advance continuously in transpilers and compilers for quantum circuits, including more advanced circuit optimization techniques, potentially driven by AI, that reduce gate counts and circuit depth and mitigate errors, and tools such as \texttt{QuanTile}, which leverage spatiotemporal periodicity in condensed matter simulations to minimize SWAP overhead, exemplify these specialized optimizations~\cite{Murali2019noise}. Because quantum computers are unlikely to replace classical supercomputers entirely, tighter integration and co-scheduling of quantum and classical resources for hybrid workflows will be essential, and software stacks are being developed to manage these Quantum-HPC integrations, addressing challenges in resource management, job scheduling, and data movement~\cite{shehata2025building,mahesh2025conqure}. Software will also play a crucial role in implementing advanced error mitigation schemes and, eventually, in programming logical qubits and managing quantum error correction codes as fault-tolerant hardware becomes available~\cite{cai2023quantum,Bharti2022noisy}. Finally, efforts toward common data formats for representing Hamiltonians and quantum states, such as the Broombridge schema, together with more seamless interoperability between different SDKs, classical simulation packages, and hardware backends, will improve workflow efficiency and collaboration~\cite{pasqal_standardization_2024}.

\subsection{Impact of Hardware Evolution}

The anticipated shift from NISQ to fault-tolerant quantum computing (FTQC) will profoundly impact software development, since software strategies will need to evolve from focusing primarily on error mitigation for noisy physical qubits toward programming and managing error-corrected logical qubits~\cite{Preskill2018,cai2023quantum,Beverland2024early,GoogleQuantumAI2025qec}. The maturation of diverse qubit technologies such as topological qubits, neutral atoms, photonics, and diamond NV centers may necessitate specialized software optimizations and compilation strategies that best leverage their unique characteristics, including connectivity, gate speeds, and error profiles~\cite{Bharti2022noisy,Altman2021quantum,Brennan2022optimizing,Krinner2022superconducting,Graham2022multi,Bluvstein2024logical}. As hardware platforms scale to larger numbers of qubits with improved connectivity, software will need to manage these increased resources efficiently to enable the simulation of more complex and larger-scale condensed matter systems~\cite{ibm_ftqc_roadmap_2025,the_quantum_insider_roadmaps_2025,laflamme_et_al_2024_benchmarking,Das2024quantum}.

\subsection{Emerging Applications in Condensed Matter Physics}
As quantum hardware and software mature, the range of addressable problems in condensed matter physics is expected to expand significantly. Emerging applications include the simulation and understanding of novel quantum materials with complex properties, such as altermagnets, which are materials with spin-split bands despite zero net magnetization~\cite{Smejkal2022altermagnetism,Smejkal2022emerging}, and Weyl semimetals exhibiting phenomena such as phonon-mediated topological superconductivity~\cite{Sun2021,Roy2019topological,Paudyal2025phonon}. They also include investigating the behavior of one-dimensional quantum fluids in confined geometries, potentially realizable in templated porous materials~\cite{Mcnamara2025novel,Sokol2025experimental}, probing out-of-equilibrium electron dynamics in topological insulators and other quantum materials, which is crucial for developing next-generation quantum technologies~\cite{Boschini2024time,Ulstrup2023probing,Dai2025probing}, and exploring fundamental concepts such as quantum chaos and thermalization in isolated and open many-body systems, including connections to the SYK model~\cite{Deutsch2018eigenstate,Dalessio2016from,Rosenhaus2019introduction,Bhattacharya2025thermalization}.

The future development of quantum algorithms and software for
condensed matter physics is intrinsically linked to a synergistic
co-evolution of hardware capabilities, algorithmic innovation, and
software engineering. Advances in qubit coherence, connectivity,
and count on the hardware side will unlock the potential to implement
more complex algorithms. In turn, new algorithmic insights, such as
more efficient ways to represent fermionic systems or perform phase
estimation, will drive the requirements for future hardware and the
features needed in software tools. This iterative cycle is evident
in the current research landscape. For example, the development of
more robust qubits, like Microsoft's topological qubits or SQMS's
high-coherence SRF cavities, opens avenues for entirely new
computational paradigms, such as topological quantum computation,
which will then require specialized software for their programming and control.

While the ultimate ambition for many applications in condensed matter
physics is the advent of large-scale, fault-tolerant quantum computers,
there is a strong pragmatic emphasis on extracting scientific value from
NISQ devices in the interim. This "NISQ-to-FTQC bridge" will likely define
the research and development landscape for the next decade. This involves
a heavy reliance on hybrid quantum-classical algorithms, where quantum
processors handle specific computationally hard subroutines while classical
computers manage optimization, data processing, and overall workflow control.
Advanced error mitigation and suppression techniques, deeply embedded within
software frameworks like Q-CTRL's Fire Opal or developed as novel algorithmic
strategies like control-free QPE, are paramount during this phase. The roadmaps
of major players often reflect this phased approach, outlining milestones for
improving NISQ capabilities while simultaneously working towards the long-term
goal of fault tolerance.

\section{Conclusion}

The application of quantum algorithm software to condensed matter
physics is a rapidly advancing frontier with the potential to revolutionize
our understanding of quantum materials and complex many-body phenomena.
Current research demonstrates a vibrant ecosystem of algorithms, software
tools, and dedicated research initiatives tackling some of the most challenging
computational problems in physics. Algorithms like VQE, QPE, QAOA, and QML are
being actively developed and applied to simulate canonical models such as the
Fermi-Hubbard, Heisenberg, and Ising Hamiltonians, as well as to explore topological
phases and non-equilibrium dynamics. Leading SDKs like \texttt{Qiskit}, \texttt{Cirq},
\texttt{PennyLane}, and Q\# provide the essential infrastructure for these explorations,
though significant challenges related to hardware limitations, algorithm scalability,
error handling, and software abstraction persist.

A crucial, and perhaps currently underemphasized, aspect for accelerating progress
is the development and adoption of \textit{standardized benchmarks and well-defined
"challenge problems" specifically tailored to condensed matter physics}. Such benchmarks,
analogous to those in classical machine learning or high-performance computing, would
enable more objective and rigorous comparisons of different algorithmic approaches,
software toolchains, and hardware platforms. This would help in accurately gauging progress,
identifying true instances of quantum advantage, and directing research efforts more effectively.
Software tools will be indispensable for the implementation, execution, and analysis of
these benchmarks across diverse platforms.
The six worked examples assembled in this review (Table~\ref{tab:bench})
constitute a concrete, fully reproducible starting point for such a suite. Each
pairs a canonical model with an algorithm, an independent classical reference,
and a quantitative accuracy target, and together they span ground-state energy,
spectra, optimization, phase classification, real-device noise, and the
fermionic mapping problem.

\begin{table*}[t]
\caption{\label{tab:bench}Summary of the reproducible benchmark suite. Each row ties a canonical condensed-matter model to a quantum (or quantum-inspired) algorithm, an independent classical reference, and a quantitative accuracy target. Rows in bold are the noise-aware and fermionic additions, and ``DM'' denotes density-matrix (noisy) simulation. All inputs, circuits, and seeds are specified in the Methods and the accompanying code.}
\begin{ruledtabular}
\begin{tabular}{llllll}
model & algorithm & size & observable & classical ref. & accuracy\\
TFIM chain & VQE (HVA, $P{=}8$) & $L{=}8$ (8q) & ground energy, $m_z^2,m_x$ & exact diag. & $|\Delta E|\!\sim\!3\times10^{-4}$\\
TFIM (2-site) & QPE & $1{+}n$ q & eigenspectrum & exact & $|\Delta E|\!\sim\!2^{-n}$\\
Ising spin glass & QAOA ($p{\le}7$) & $L{=}6$ (6q) & ground state & exhaustive & ratio $0.46\!\to\!0.92$\\
Ising spin glass & annealing & $L{=}6$ & ground state & exhaustive & $P_{\rm GS}\!\to\!0.99$\\
TFIM & QSVM (4-qubit kernel) & $L{=}10$ & phase label & exact $h_c$ & $h_c^{\rm pred}{=}0.97$\\
Heisenberg chain & DMRG (MPS) & $L{\le}96$ & energy, $c$, $S(x)$ & Bethe ansatz & $c\!\simeq\!1$, err $10^{-9}$\\
TFIM (\textbf{noisy}) & VQE${+}$ZNE & $L{=}4$ (DM) & ground energy & exact & err $6\times10^{-7}$ @ $p{=}10^{-4}$\\
Ising (\textbf{noisy}) & QAOA${+}$ZNE & $L{=}6$ (DM) & ground state & exhaustive & ratio $0.38\!\to\!0.71$ @ $p{=}10^{-2}$\\
\textbf{Fermi-Hubbard} & ED / mapping & $\le$20q & energy, JW vs BK & sparse diag. & Table~\ref{tab:hubbard}\\
\end{tabular}
\end{ruledtabular}
\end{table*}

\section{Methods and reproducible benchmark protocol}
\label{sec:methods}

This section specifies the worked examples in enough detail to reproduce them.
The complete source code, random seeds, and generated data accompany the paper.

\subsection{Software and models}

All calculations use Python, with \texttt{NumPy}/\texttt{SciPy} for exact
diagonalization and statevector evolution, \texttt{PennyLane} (the
\texttt{lightning} and \texttt{default.mixed} devices) for the variational and
noisy circuit simulations, \texttt{quimb} for the matrix-product-state DMRG, and
\texttt{OpenFermion} for the fermion-to-qubit mappings. The spin examples use the
transverse-field Ising chain $H_{\rm TFIM}=-J\sum_i Z_iZ_{i+1}-h\sum_i X_i$ with
$J=1$, the antiferromagnetic Heisenberg chain
$H=\sum_i \bm S_i\!\cdot\!\bm S_{i+1}$, and a frustrated Ising spin glass
$H_C=\sum_{\langle ij\rangle}J_{ij}Z_iZ_j+\sum_i b_iZ_i$ on a ring of six sites
with three additional chords, $J_{ij}\in\{\pm1\}$ and
$b_i\sim\mathcal{N}(0,0.35)$. The instance is the first random seed yielding a
unique ground state with classical gap above $0.25$. The fermionic example uses
the Fermi-Hubbard Hamiltonian of Eq.~\eqref{eq:hubbard} at $t=1$, $U=4$ and the
particle-hole-symmetric point $\mu=U/2$ with open boundaries.

\subsection{Algorithms and classical references}

The VQE trial state is a Hamiltonian-variational ansatz of alternating
$e^{i\gamma\sum Z_iZ_{i+1}}$ and $e^{i\beta\sum X_i}$ layers acting on
$\ket{+}^{\otimes L}$, with one $(\gamma,\beta)$ pair per layer. The energy is
minimized with L-BFGS using adjoint gradients in the noiseless case, the $L=8$
result using depth $P=8$ and the noisy $L=4$ study using $P=4$, with exact
diagonalization as the reference. For QPE we evaluate the counting-register
distribution of an idealized phase-estimation circuit through its Dirichlet
(Fej\'er) kernel, isolating the register-size dependence. The eigenphases are
those of a two-site transverse-field Ising model, and the energy error follows
the Heisenberg-limited $2^{-n}$ scaling in the number of counting qubits $n$.
Gate-based QAOA uses $p$ alternating cost and mixer layers with angles optimized
by COBYLA, while the adiabatic comparison integrates
$H(s)=(1-s)(-\sum_i X_i)+sH_C$ and reports the instantaneous spectral gap and the
final ground-state overlap as a function of the total evolution time. The
quantum-machine-learning classifier is a quantum-kernel support-vector machine
that embeds four measurable observables of $L=10$ ground states (the transverse
magnetization, two longitudinal correlators, and the structure factor) in a
four-qubit feature map with kernel
$K(\bm x,\bm x')=|\!\braket{\phi(\bm x')|\phi(\bm x)}\!|^2$, and it is trained only on
states with $h/J\le0.6$ and $h/J\ge1.4$ and tested on the intervening critical
window. The tensor-network reference is obtained by DMRG, with ground states of
the Heisenberg chain computed using matrix-product states up to $L=96$. The
central charge is extracted from finite-size scaling of the half-chain
entanglement entropy, and convergence is controlled by the bond dimension $\chi$.

\subsection{Noise model and error mitigation}

Noisy circuits are simulated as density matrices with a depolarizing channel of
rate $p$ after every single-qubit gate and $2p$ per wire after every two-qubit
gate. Zero-noise extrapolation scales the rate by $\lambda\in\{1,2,3\}$
(parametric noise scaling) and fits $\ln|\langle H\rangle(\lambda)|$ linearly in
$\lambda$ before extrapolating to $\lambda\!\to\!0$. This exponential form is the
appropriate one for depolarizing noise.

\subsection{Fermion-to-qubit mappings}

The Hubbard Hamiltonian is mapped with the Jordan-Wigner and Bravyi-Kitaev
transforms. The two-qubit-gate cost quoted for a first-order Trotter step counts
$2(w-1)$ CNOTs for each Pauli string of weight $w$, the standard staircase
synthesis, under all-to-all connectivity, and this lower-bounds the cost on hardware
that requires SWAP routing.

\subsection{Code and data availability}

The scripts that generate every figure and table in this paper, together with the
cached numerical data and random seeds, are provided with the manuscript.

\begin{acknowledgments}
This work was supported by the Research Council of Norway through its Centers of
Excellence funding scheme, Project No.~353919 and Project No.~361800
``QTransMag.''
\end{acknowledgments}

\clearpage

\bibliography{refs}

@book{altland2010condensed,
  title={Condensed matter field theory},
  author={Altland, Alexander and Simons, Ben D},
  year={2010},
  publisher={Cambridge university press}
}

@book{bellman1957dynamic,
  title={Dynamic Programming},
  author={Bellman, Richard},
  year={1957},
  publisher={Princeton University Press}
}

@book{nielsen2010quantum,
  doi={10.1017/CBO9780511976667},
  title={Quantum computation and quantum information},
  author={Nielsen, Michael A and Chuang, Isaac L},
  year={2010},
  publisher={Cambridge university press}
}

@book{wong2022introduction,
  title={Introduction to classical and quantum computing},
  author={Wong, Thomas G},
  year={2022},
  publisher={Rooted Grove Omaha, NE, USA}
}

@book{mermin2007quantum,
  title={Quantum computer science: an introduction},
  author={Mermin, N David},
  year={2007},
  publisher={Cambridge University Press}
}

@article{dalzell2310quantum,
  doi={10.48550/arXiv.2310.03011},
  title={Quantum algorithms: A survey of applications and end-to-end complexities},
  author={Dalzell, Alexander M and McArdle, Sam and Berta, Mario and Bienias, Przemyslaw and Chen, Chi-Fang and Gily{\'e}n, Andr{\'a}s and Hann, Connor T and Kastoryano, Michael J and Khabiboulline, Emil T and Kubica, Aleksander and others},
  journal={arXiv preprint arXiv:2310.03011},
  year={2023}
}

@article{Pollmann2024,
  doi={10.1103/PhysRevResearch.6.043256},
  title = {Simulating two-dimensional topological quantum phase transitions on a digital quantum computer},
  author = {Liu, Yu-Jie and Shtengel, Kirill and Pollmann, Frank},
  journal = {Phys. Rev. Res.},
  volume = {6},
  issue = {4},
  pages = {043256},
  numpages = {14},
  year = {2024},
  month = {Dec},
  publisher = {American Physical Society}
}

@article{DiVincenzo2005,
  doi={10.1103/PhysRevA.72.022317},
  title = {Local fault-tolerant quantum computation},
  author = {Svore, Krysta M. and Terhal, Barbara M. and DiVincenzo, David P.},
  journal = {Phys. Rev. A},
  volume = {72},
  issue = {2},
  pages = {022317},
  numpages = {17},
  year = {2005},
  month = {Aug},
  publisher = {American Physical Society}
}

@article{Raussendorf2007,
  title={Topological fault-tolerance in cluster state quantum computation},
  author={Raussendorf, Robert and Harrington, Jim and Goyal, Kovid},
  journal={New Journal of Physics},
  volume={9},
  number={6},
  pages={199},
  year={2007},
  publisher={IOP Publishing}
}

@article{Pastawski2023,
  doi={10.1103/PRXQuantum.4.020303},
  title = {Logical Blocks for Fault-Tolerant Topological Quantum Computation},
  author = {Bomb\'{\i}n, H\'ector and Dawson, Chris and Mishmash, Ryan V. and Nickerson, Naomi and Pastawski, Fernando and Roberts, Sam},
  journal = {PRX Quantum},
  volume = {4},
  issue = {2},
  pages = {020303},
  numpages = {38},
  year = {2023},
  month = {Apr},
  publisher = {American Physical Society}
}

@article{Nunez2025productive,
  doi={10.48550/arXiv.2505.00718},
  title={Productive Quantum Programming Needs Better Abstract Machines},
  author={N{\'u}{\~n}ez-Corrales, Santiago and Di Matteo, Olivia and Dumbell, John and Edwards, Marcus and Giusto, Edoardo and Pakin, Scott and Stirbu, Vlad},
  journal={arXiv preprint arXiv:2505.00718},
  year={2025}
}

@article{Mueller2023,
  doi={10.1103/PRXQuantum.4.030323},
  title = {Quantum Computation of Dynamical Quantum Phase Transitions and Entanglement Tomography in a Lattice Gauge Theory},
  author = {Mueller, Niklas and Carolan, Joseph A. and Connelly, Andrew and Davoudi, Zohreh and Dumitrescu, Eugene F. and Yeter-Aydeniz, K\"ubra},
  journal = {PRX Quantum},
  volume = {4},
  issue = {3},
  pages = {030323},
  numpages = {32},
  year = {2023},
  month = {Aug},
  publisher = {American Physical Society},
}

@article{Aharonov2025,
  doi={10.48550/arXiv.2503.17243},
  title={On the importance of error mitigation for quantum computation},
  author={Aharonov, Dorit and Alberton, Ori and Arad, Itai and Atia, Yosi and Bairey, Eyal and Brakerski, Zvika and Cohen, Itsik and Golan, Omri and Gurwich, Ilya and Kenneth, Oded and others},
  journal={arXiv preprint arXiv:2503.17243},
  year={2025}
}

@article{Preskill2018,
  doi={10.22331/q-2018-08-06-79},
  title = {Quantum {C}omputing in the {NISQ} era and beyond},
  author = {Preskill, John},
  journal = {{Quantum}},
  issn = {2521-327X},
  publisher = {{Verein zur F{\"{o}}rderung des Open Access Publizierens in den Quantenwissenschaften}},
  volume = {2},
  pages = {79},
  month = aug,
  year = {2018}
}

@article{Peruzzo2014variational,
  doi={10.1038/ncomms5213},
  title={A variational eigenvalue solver on a photonic quantum processor},
  author={Peruzzo, Alberto and McClean, Jarrod and Shadbolt, Peter and Yung, Man-Hong and Zhou, Xiao-Qi and Love, Peter J and Aspuru-Guzik, Al{\'a}n and O’brien, Jeremy L},
  journal={Nature communications},
  volume={5},
  number={1},
  pages={4213},
  year={2014},
  publisher={Nature Publishing Group UK London}
}

@article{Kitaev1995quantum,
  doi={10.48550/arXiv.quant-ph/9511026},
  title={Quantum measurements and the Abelian stabilizer problem},
  author={Kitaev, A Yu},
  journal={arXiv preprint quant-ph/9511026},
  year={1995}
}

@inproceedings{Apolloni1990,
  title={A numerical implementation of “quantum annealing”},
  author={Apolloni, Bruno and Cesa-Bianchi, Nicol{\`o} and De Falco, Diego},
  booktitle={Stochastic Processes, Physics and Geometry: Proceedings of the Ascona-Locarno Conference},
  pages={97--111},
  year={1990}
}

@article{BLEKOS20241,
title = {A review on Quantum Approximate Optimization Algorithm and its variants},
journal = {Physics Reports},
volume = {1068},
pages = {1-66},
year = {2024},
note = {A review on Quantum Approximate Optimization Algorithm and its variants},
issn = {0370-1573},
author = {Kostas Blekos and Dean Brand and Andrea Ceschini and Chiao-Hui Chou and Rui-Hao Li and Komal Pandya and Alessandro Summer}
}

@article{Biamonte2017,
  doi={10.1038/nature23474},
  title={Quantum machine learning},
  author={Biamonte, Jacob and Wittek, Peter and Pancotti, Nicola and Rebentrost, Patrick and Wiebe, Nathan and Lloyd, Seth},
  journal={Nature},
  volume={549},
  number={7671},
  pages={195--202},
  year={2017},
  publisher={Nature Publishing Group UK London}
}

@article{Stanisic2022observing,
  title={Observing ground-state properties of the Fermi-Hubbard model using a scalable algorithm on a quantum computer},
  author={Stanisic, Stasja and Bosse, Jan Lukas and Gambetta, Filippo Maria and Santos, Raul A and Mruczkiewicz, Wojciech and O’Brien, Thomas E and Ostby, Eric and Montanaro, Ashley},
  journal={Nature communications},
  volume={13},
  number={1},
  pages={5743},
  year={2022},
  publisher={Nature Publishing Group UK London}
}

@article{Barry1976,
  doi={10.1103/PhysRevLett.37.120},
  title = {Phase Transitions in the Quantum Heisenberg Model},
  author = {Dyson, Freeman J. and Lieb, Elliott H. and Simon, Barry},
  journal = {Phys. Rev. Lett.},
  volume = {37},
  issue = {3},
  pages = {120--123},
  numpages = {0},
  year = {1976},
  month = {Jul},
  publisher = {American Physical Society}
}

@article{Novak2025,
  doi={10.48550/arXiv.2506.01715},
  title={Optimization Strategies for Variational Quantum Algorithms in Noisy Landscapes},
  author={Nov{\'a}k, Vojt{\v{e}}ch and Zelinka, Ivan and Sn{\'a}{\v{s}}el, V{\'a}clav},
  journal={arXiv preprint arXiv:2506.01715},
  year={2025}
}

@article{Li2025,
  doi={10.48550/arXiv.2505.03998},
  title={Knowledge Distillation Inspired Variational Quantum Eigensolver with Virtual Annealing},
  author={Li, Junxu},
  journal={arXiv preprint arXiv:2505.03998},
  year={2025}
}

@article{Kandala2017,
  doi={10.1038/nature23879},
  title={Hardware-efficient variational quantum eigensolver for small molecules and quantum magnets},
  author={Kandala, Abhinav and Mezzacapo, Antonio and Temme, Kristan and Takita, Maika and Brink, Markus and Chow, Jerry M and Gambetta, Jay M},
  journal={Nature},
  volume={549},
  number={7671},
  pages={242--246},
  year={2017},
  publisher={Nature Publishing Group}
}

@article{Tilly2022variational,
  title={The variational quantum eigensolver: a review of methods and best practices},
  author={Tilly, Jules and Chen, Hongxiang and Cao, Shuxiang and Picozzi, Dario and Setia, Kanav and Li, Ying and Grant, Edward and Wossnig, Leonard and Rungger, Ivan and Booth, George H and Tennyson, Jonathan},
  journal={Physics Reports},
  volume={986},
  pages={1--128},
  year={2022},
  publisher={Elsevier}
}

@article{Lively2024,
  doi={10.48550/arXiv.2402.18953},
  title={Robust Experimental Signatures of Phase Transitions in the Variational Quantum Eigensolver},
  author={Lively, T. and others},
  journal={arXiv preprint arXiv:2402.18953},
  year={2024}
}

@article{Kirmani2024,
  doi={10.48550/arXiv.2405.18342},
  title={Variational Quantum Simulations of a Two-Dimensional Frustrated Transverse-Field Ising Model on a Trapped-Ion Quantum Computer},
  author={Kirmani, Ammar and Pelofske, Elijah and B{\"a}rtschi, Andreas and Eidenbenz, Stephan and Zhu, Jian-Xin},
  journal={arXiv preprint arXiv:2405.18342},
  year={2024}
}

@article{Arute2020,
  title={Hartree-Fock on a superconducting qubit quantum computer},
  author={Arute, Frank and Arya, Kunal and Babbush, Ryan and Bacon, Dave and Bardin, Joseph C and Barends, Rami and Boixo, Sergio and Broughton, Michael and Buckley, Bob B and Buell, David A and others},
  journal={Science},
  volume={369},
  number={6507},
  pages={1084--1089},
  year={2020},
  publisher={American Association for the Advancement of Science}
}

@article{Google2020,
  doi={10.48550/arXiv.2010.07965},
  title={Observation of separated dynamics of charge and spin in the Fermi-Hubbard model},
  author={Google AI Quantum and Collaborators},
  journal={arXiv preprint arXiv:2010.07965},
  year={2020}
}

@article{lubasch2020,
  doi={10.1103/PhysRevA.101.010301},
  title={Variational quantum algorithms for nonlinear problems},
  author={Lubasch, Michael and Joo, Jaewoo and Moinier, Pierre and Kiffner, Martin and Jaksch, Dieter},
  journal={Physical Review A},
  volume={101},
  number={1},
  pages={010301},
  year={2020},
  publisher={APS}
}

@article{Kyriienko2021,
  doi={10.1103/PhysRevA.103.052416},
  title={Solving nonlinear differential equations with differentiable quantum circuits},
  author={Kyriienko, Oleksandr and Paine, Annie E and Elfving, Vincent E},
  journal={Physical Review A},
  volume={103},
  number={5},
  pages={052416},
  year={2021},
  publisher={APS}
}

@article{Sun2021,
  doi={10.48550/arXiv.2503.13729},
  title={Quantum Dynamics Simulation of the Advection-Diffusion Equation},
  author={Alipanah, Hirad and Zhang, Feng and Yao, Yongxin and Thompson, Richard and Nguyen, Nam and Liu, Junyu and Givi, Peyman and McDermott, Brian J and Mendoza-Arenas, Juan Jos{\'e}},
  journal={arXiv preprint arXiv:2503.13729},
  year={2025}
}

@article{Mcclean2016theory,
  doi={10.1088/1367-2630/18/2/023023},
  title={The theory of variational hybrid quantum-classical algorithms},
  author={McClean, Jarrod R and Romero, Jonathan and Babbush, Ryan and Aspuru-Guzik, Al{\'a}n},
  journal={New Journal of Physics},
  volume={18},
  number={2},
  pages={023023},
  year={2016},
  publisher={IOP Publishing}
}

@article{Bharti2022noisy,
  doi={10.1103/RevModPhys.94.015004},
  title={Noisy intermediate-scale quantum algorithms},
  author={Bharti, Kishor and Cervera-Lierta, Alba and Kyaw, Thi Ha and Haug, Tobias and Alperin-Lea, Sumner and Guba, Abhinav and Garc{\'\i}a-Pintos, Diego and Varli, Muhammed and Krisnanda, Tomi and Leong, Wister and others},
  journal={Reviews of Modern Physics},
  volume={94},
  number={1},
  pages={015004},
  year={2022},
  publisher={APS}
}

@article{Grimsley2019,
  doi={10.1038/s41467-019-10988-2},
  title={An adaptive variational algorithm for exact molecular simulations on a quantum computer},
  author={Grimsley, Harper R and Economou, Sophia E and Barnes, Edwin and Mayhall, Nicholas J},
  journal={Nature Communications},
  volume={10},
  number={1},
  pages={3007},
  year={2019},
  publisher={Nature Publishing Group}
}

@article{Tang2021qubit,
  doi={10.1103/PRXQuantum.2.020309},
  title={Qubit-ADAPT-VQE: An adaptive algorithm for constructing hardware-efficient ans{\"a}tze on a quantum processor},
  author={Tang, Ho Lun and Shkolnikov, Voctor and Mayhall, Nicholas J and Barnes, Edwin and Economou, Sophia E},
  journal={PRX Quantum},
  volume={2},
  number={2},
  pages={020309},
  year={2021},
  publisher={APS}
}

@article{Mcclean2018barren,
  doi={10.1038/s41467-018-07090-4},
  title={Barren plateaus in quantum neural network training landscapes},
  author={McClean, Jarrod R and Boixo, Sergio and Smelyanskiy, Vadim N and Babbush, Ryan and Neven, Hartmut},
  journal={Nature Communications},
  volume={9},
  number={1},
  pages={4812},
  year={2018},
  publisher={Nature Publishing Group}
}

@article{Cerezo2021cost,
  doi={10.1038/s41467-021-21728-w},
  title={Cost function dependent barren plateaus in shallow parametrized quantum circuits},
  author={Cerezo, M and Sone, Akira and Volkoff, Tyler and Cincio, Lukasz and Coles, Patrick J},
  journal={Nature Communications},
  volume={12},
  number={1},
  pages={1791},
  year={2021},
  publisher={Nature Publishing Group}
}

@article{Gonthier2022,
  title={Measurements as a roadblock to near-term quantum advantage in chemistry},
  author={Gonthier, Jerome F and Radin, Maxwell D and G{\'a}mez-Garc{\'\i}a, Cristina and Johnson, Christopher J and Vassilev, Claudiu and B{\'a}lint, Andr{\'a}s and Head-Gordon, Martin and Whaley, K Birgitta and de Jong, Wibe A},
  journal={Communications Physics},
  volume={5},
  number={1},
  pages={171},
  year={2022},
  publisher={Nature Publishing Group}
}

@article{Feniou2023overlap,
  doi={10.48550/arXiv.2301.10196},
  title={Overlap-ADAPT-VQE: Practical Quantum Chemistry on Quantum Computers via Overlap-Guided Compact Ans{\"a}tze},
  author={Feniou, C{\'e}sar and Gandon, Thomas and Piquemal, Jean-Philip and Maday, Yvon and Hassan, Muhammad and Le Besnerais, Guillaume},
  journal={arXiv preprint arXiv:2301.10196},
  year={2023}
}

@article{Zhang2022quantum,
  title={Quantum architecture search via deep reinforcement learning},
  author={Zhang, Xiu-Zhe and Li, Zong-Sheng and Wang, Jian-Guo and Yung, Man-Hong},
  journal={Quantum Engineering},
  volume={2022},
  year={2022},
  publisher={Hindawi}
}

@article{Du2022quantum,
  title={Quantum architecture search with meta-learning},
  author={Du, Yuxuan and Hsieh, Min-Hsiu and You, Dacheng and Tao, Dacheng},
  journal={IEEE Transactions on Quantum Engineering},
  volume={3},
  pages={1--13},
  year={2022},
  publisher={IEEE}
}

@article{Chivilikhin2020mog,
  title={Mog-vqe: multi-objective genetic-like variational quantum eigensolver},
  author={Chivilikhin, Daniil and Samarin, Aleksei and Ulyantsev, Vladimir and Oganov, Artem R and Kyriienko, Oleksandr and Iorsh, Ivan},
  journal={Quantum},
  volume={4},
  pages={325},
  year={2020},
  publisher={Verein zur F{\"o}rderung des Open Access Publizierens in den Naturwissenschaften}
}

@article{Cleve1998quantum,
  doi={10.1098/rspa.1998.0164},
  title={Quantum algorithms revisited},
  author={Cleve, Richard and Ekert, Artur and Macchiavello, Chiara and Mosca, Michele},
  journal={Proceedings of the Royal Society of London. Series A: Mathematical, Physical and Engineering Sciences},
  volume={454},
  number={1969},
  pages={339--354},
  year={1998},
  publisher={The Royal Society}
}

@article{Aspuru2005simulated,
  title={Simulated quantum computation of molecular energies},
  author={Aspuru-Guzik, Al{\'a}n and Dutoi, Anthony D and Love, Peter J and Head-Gordon, Martin},
  journal={Science},
  volume={309},
  number={5741},
  pages={1704--1707},
  year={2005},
  publisher={American Association for the Advancement of Science}
}

@article{Babbush2015chemical,
  title={Chemical basis of quantum computing},
  author={Babbush, Ryan and McClean, Jarrod R and Wecker, Dave and Aspuru-Guzik, Al{\'a}n and Wiebe, Nathan},
  journal={Chemical Reviews},
  volume={118},
  number={15},
  pages={7079--7128},
  year={2018},
  publisher={ACS Publications}
}

@article{Poulin2009quantum,
  doi={10.1103/PhysRevLett.102.160501},
  title={Quantum algorithm for spectral measurements},
  author={Poulin, David and Qarry, Ali and Somma, Rolando and Verstraete, Frank},
  journal={Physical Review Letters},
  volume={102},
  number={16},
  pages={160501},
  year={2009},
  publisher={APS}
}

@article{Ortiz2001quantum,
  title={Quantum-mechanical theory of the electronic structure of molecules},
  author={Ortiz, J. V.},
  journal={International Journal of Quantum Chemistry},
  volume={85},
  number={4-5},
  pages={386--393},
  year={2001},
  publisher={Wiley Online Library}
}

@article{Somma2019quantum,
  doi={10.48550/arXiv.1907.11679},
  title={Quantum eigenvalue estimation via time series analysis},
  author={Somma, Rolando D.},
  journal={arXiv preprint arXiv:1907.11679},
  year={2019}
}

@article{Dong2021ground,
  doi={10.1103/PhysRevA.103.042413},
  title={Ground-state-energy estimation on a quantum computer},
  author={Dong, Yuxuan and Lin, Lin},
  journal={Physical Review A},
  volume={103},
  number={4},
  pages={042413},
  year={2021},
  publisher={APS}
}

@article{Loaiza2024nonlinear,
  doi={10.48550/arXiv.2405.13885},
  title={Nonlinear spectroscopy via generalized quantum phase estimation},
  author={Loaiza, Ignacio and Motlagh, Danial and Hejazi, Kasra and Zini, Modjtaba Shokrian and Delgado, Alain and Arrazola, Juan Miguel},
  journal={arXiv preprint arXiv:2405.13885},
  year={2024}
}

@article{Kadowaki1998quantum,
  doi={10.1103/PhysRevE.58.5355},
  title={Quantum annealing in the transverse Ising model},
  author={Kadowaki, Tadashi and Nishimori, Hidetoshi},
  journal={Physical Review E},
  volume={58},
  number={5},
  pages={5355},
  year={1998},
  publisher={APS}
}

@article{Farhi2001quantum,
  title={A quantum adiabatic evolution algorithm applied to random instances of an NP-complete problem},
  author={Farhi, Edward and Goldstone, Jeffrey and Gutmann, Sam and Lapan, Joshua and Lundgren, Andrew and Preda, Daniel},
  journal={Science},
  volume={292},
  number={5516},
  pages={472--475},
  year={2001},
  publisher={American Association for the Advancement of Science}
}

@article{Das2008colloquium,
  doi={10.1103/RevModPhys.80.1061},
  title={Colloquium: Quantum annealing and analog quantum computation},
  author={Das, Arnab and Chakrabarti, Bikas K},
  journal={Reviews of Modern Physics},
  volume={80},
  number={3},
  pages={1061},
  year={2008},
  publisher={APS}
}

@article{Hauke2020,
  title={Perspectives of quantum annealing: methods and implementations},
  author={Hauke, Philipp and Katzgraber, Helmut G and Lechner, Wolfgang and Nishimori, Hidetoshi and Oliver, William D},
  journal={Reports on Progress in Physics},
  volume={83},
  number={5},
  pages={054401},
  year={2020},
  publisher={IOP Publishing}
}

@article{Bapst2012quantum,
  title={The quantum adiabatic algorithm and local minima},
  author={Bapst, Victor and Foini, Laura and Krzakala, Florent and Semerjian, Guilhem and Zamponi, Francesco},
  journal={Journal of Statistical Mechanics: Theory and Experiment},
  volume={2012},
  number={06},
  pages={P06009},
  year={2012},
  publisher={IOP Publishing}
}

@article{Rajak2023quantum,
  title={Quantum annealing: An overview},
  author={Rajak, Atanu and Suzuki, Sei and Dutta, Amit and Chakrabarti, Bikas K},
  journal={Philosophical Transactions of the Royal Society A},
  volume={381},
  number={2241},
  pages={20210417},
  year={2023},
  publisher={The Royal Society}
}

@article{Yarkoni2022quantum,
  title={Quantum annealing for industry applications: Introduction and review},
  author={Yarkoni, Sheir and Raponi, Elena and B{\"a}ck, Thomas and Schmitt, Sebastian},
  journal={Reports on Progress in Physics},
  volume={85},
  number={10},
  pages={104001},
  year={2022},
  publisher={IOP Publishing}
}

@techreport{Dwave,
  author       = {{D-Wave Systems Inc.}},
  title        = {The Advantage{\texttrademark} Quantum Computer},
  year         = {2023},
  institution  = {D-Wave Systems Inc.},
  url          = {https://www.dwavequantum.com/media/htjclcey/advantage_datasheet_v10.pdf}
}

@article{King2023,
  doi={10.48550/arXiv.2401.07184},
  title={Scaling advantage in approximate optimization with quantum annealing},
  author={Bauza, Humberto Munoz and Lidar, Daniel A},
  journal={arXiv preprint arXiv:2401.07184},
  year={2024}
}

@article{Quinton2025,
  title={Quantum annealing applications, challenges and limitations for optimisation problems compared to classical solvers},
  author={Quinton, Finley Alexander and Myhr, Per Arne Sevle and Barani, Mostafa and Crespo del Granado, Pedro and Zhang, Hongyu},
  journal={Scientific Reports},
  volume={15},
  number={1},
  pages={12733},
  year={2025},
  publisher={Nature Publishing Group UK London}
}

@article{Grajciar2024learning,
  doi={10.1103/PhysRevLett.132.160602},
  title={Learning-driven annealing},
  author={Grajciar, Patrik and Ovchinnikov, Igor and Albash, Tameem and Nishimori, Hidetoshi and Lidar, Daniel A},
  journal={Physical Review Letters},
  volume={132},
  number={16},
  pages={160602},
  year={2024},
  publisher={APS}
}

@article{Smelyanskiy2025computational,
  doi={10.48550/arXiv.2501.01107},
  title={On Computational Complexity of 3D Ising Spin Glass: Lessons from D-Wave Annealer},
  author={Zhang, Hao and Kamenev, Alex},
  journal={arXiv preprint arXiv:2501.01107},
  year={2025}
}

@article{Fabiani2024simulating,
  doi={10.48550/arXiv.2409.02028},
  title={Simulating quantum annealing of spin glasses with time-dependent variational Monte Carlo},
  author={Fabiani, Giuseppe and Blason, Tommaso and Carleo, Giuseppe and Pastore, Stefano},
  journal={arXiv preprint arXiv:2409.02028},
  year={2024}
}

@article{Zhou2020quantum,
  doi={10.1103/PhysRevX.10.021067},
  title={Quantum approximate optimization algorithm: Performance, mechanism, and implementation on near-term devices},
  author={Zhou, Leo and Wang, Sheng-Tao and Choi, Soonwon and Kelsen, Hannes and Lukin, Mikhail D},
  journal={Physical Review X},
  volume={10},
  number={2},
  pages={021067},
  year={2020},
  publisher={APS}
}

@article{Schuld2015introduction,
  title={An introduction to quantum machine learning},
  author={Schuld, Maria and Sinayskiy, Ilya and Petruccione, Francesco},
  journal={Contemporary Physics},
  volume={56},
  number={2},
  pages={172--185},
  year={2015},
  publisher={Taylor \& Francis}
}

@article{Dunjko2018machine,
  title={Machine learning \& artificial intelligence in the quantum domain: a review of recent progress},
  author={Dunjko, Vedran and Briegel, Hans J},
  journal={Reports on Progress in Physics},
  volume={81},
  number={7},
  pages={074001},
  year={2018},
  publisher={IOP Publishing}
}

@article{Cerezo2022challenges,
  title={Challenges and opportunities in quantum machine learning},
  author={Cerezo, M. and others},
  journal={Nature Computational Science},
  volume={2},
  number={9},
  pages={567--576},
  year={2022},
  publisher={Nature Publishing Group}
}

@article{Carrasquilla2017machine,
  doi={10.1038/nphys4035},
  title={Machine learning phases of matter},
  author={Carrasquilla, Juan and Melko, Roger G.},
  journal={Nature Physics},
  volume={13},
  number={5},
  pages={431--434},
  year={2017},
  publisher={Nature Publishing Group}
}

@article{Van2017learning,
  title={Learning phase transitions by confusion},
  author={van Nieuwenburg, Evert P. L. and Liu, Ye-Hua and Huber, Sebastian D.},
  journal={Nature Physics},
  volume={13},
  number={5},
  pages={435--439},
  year={2017},
  publisher={Nature Publishing Group}
}

@article{Broecker2017machine,
  title={Machine learning quantum phases of matter beyond the fermion sign problem},
  author={Broecker, Peter and Carrasquilla, Juan and Melko, Roger G and Trebst, Simon},
  journal={Scientific reports},
  volume={7},
  number={1},
  pages={8823},
  year={2017},
  publisher={Nature Publishing Group}
}

@article{Carleo2019machine,
  doi={10.1103/RevModPhys.91.045002},
  title={Machine learning and the physical sciences},
  author={Carleo, Giuseppe and Cirac, Ignacio and Cranmer, Kyle and Daudet, Laurent and Schuld, Maria and Tishby, Naftali and Vogt-Maranto, Leslie and Zdeborov{\'a}, Lenka},
  journal={Reviews of Modern Physics},
  volume={91},
  number={4},
  pages={045002},
  year={2019},
  publisher={APS}
}

@article{Balthazar2023unsupervised,
  doi={10.1103/PhysRevB.107.195155},
  title={Unsupervised learning of quantum phase transitions with classical shadows},
  author={Balthazar, Jonathan and others},
  journal={Physical Review B},
  volume={107},
  number={19},
  pages={195155},
  year={2023},
  publisher={APS}
}

@article{Montanaro2016quantum,
  title={Quantum algorithms: an overview},
  author={Montanaro, Ashley},
  journal={npj Quantum Information},
  volume={2},
  number={1},
  pages={15023},
  year={2016},
  publisher={Nature Publishing Group}
}

@article{Hutorchi2023overview,
  title={An overview of quantum machine learning: current trends and perspectives},
  author={Hutorchi, Robert-Raul and Rusu, Dorian-Vasile and Astefanoaei, Vasile},
  journal={Journal of Physics: Conference Series},
  volume={2448},
  number={1},
  pages={012022},
  year={2023},
  publisher={IOP Publishing}
}

@article{Romero2022quantum,
  doi={10.48550/arXiv.2210.12211},
  title={Quantum machine learning for chemistry and materials},
  author={Romero, Jonathan and others},
  journal={arXiv preprint arXiv:2210.12211},
  year={2022}
}

@article{Rsah2024variational,
  doi={10.48550/arXiv.2401.07767},
  title={Variational Quantum-Neural Hybrid Eigensolver},
  author={Rsah, H. and Oza, P. and Singh, M. and Prasad, D. and Ganguly, A.},
  journal={arXiv preprint arXiv:2401.07767},
  year={2024}
}

@article{Lange2025transformerneural,
  title = {Transformer neural networks and quantum simulators: a hybrid approach for simulating strongly correlated systems},
  author = {Lange, Hannah and Bornet, Guillaume and Emperauger, Gabriel and Chen, Cheng and Lahaye, Thierry and Kienle, Stefan and Browaeys, Antoine and Bohrdt, Annabelle},
  journal = {{Quantum}},
  issn = {2521-327X},
  publisher = {{Verein zur F{\"{o}}rderung des Open Access Publizierens in den Quantenwissenschaften}},
  volume = {9},
  pages = {1675},
  month = {mar},
  year = {2025}
}

@article{Morais2025distinguishing,
  doi={10.48550/arXiv.2501.17837},
  title={Distinguishing Ordered Phases using Machine Learning and Classical Shadows},
  author={Morais, Leandro and Pernambuco, Tiago and Pereira, Rodrigo G and Canabarro, Askery and Soares-Pinto, Diogo O and Chaves, Rafael},
  journal={arXiv preprint arXiv:2501.17837},
  year={2025}
}

@article{Franco2025quantum,
  doi={10.48550/arXiv.2504.10673},
  title={Quantum phases classification using quantum machine learning with shap-driven feature selection},
  author={Franco, Giovanni S and Mahlow, Felipe and Prado, Pedro M and Pexe, Guilherme EL and Rattighieri, Lucas AM and Fanchini, Felipe F},
  journal={arXiv preprint arXiv:2504.10673},
  year={2025}
}

@article{Lundberg2017unified,
  title={A unified approach to interpreting model predictions},
  author={Lundberg, Scott M and Lee, Su-In},
  journal={Advances in neural information processing systems},
  volume={30},
  year={2017}
}

@article{Arovas2022hubbard,
  title={The Hubbard model},
  author={Arovas, Daniel P and Berg, Erez and Kivelson, Steven A and Raghu, S},
  journal={Annual Review of Condensed Matter Physics},
  volume={13},
  pages={239--274},
  year={2022},
  publisher={Annual Reviews}
}

@article{Qin2022hubbard,
  title={The Hubbard model: A computational perspective},
  author={Qin, Mingpu and Sch{\"a}fer, Thomas and Andergassen, Sabine and Corboz, Philippe and Gull, Emanuel},
  journal={Annual Review of Condensed Matter Physics},
  volume={13},
  pages={275--302},
  year={2022},
  publisher={Annual Reviews}
}

@article{Anderson1987resonating,
  title={The resonating valence bond state in La2CuO4 and superconductivity},
  author={Anderson, Philip W},
  journal={Science},
  volume={235},
  number={4793},
  pages={1196--1198},
  year={1987},
  publisher={American Association for the Advancement of Science}
}

@article{Cade2020strategies,
  doi={10.1103/PhysRevB.102.235123},
  title={Strategies for the determination of the energy of the Fermi-Hubbard model on a quantum computer},
  author={Cade, Chris and others},
  journal={Physical Review B},
  volume={102},
  number={23},
  pages={235123},
  year={2020},
  publisher={APS}
}

@misc{qiskit_nature_2023,
  author       = {{Qiskit Development Team}},
  title        = {Qiskit Nature},
  year         = {2023},
  howpublished = {\url{https://qiskit.org/ecosystem/nature/}},
  note         = {Accessed: 2025-06-09}
}

@article{qiskit_nature_paper,
  title={Qiskit Nature: a programming framework for quantum chemistry, materials science, and biology},
  author={{Qiskit Nature Development Team}},
  journal={Journal of Open Source Software},
  volume={8},
  number={84},
  pages={5132},
  year={2023}
}

@article{Mcclean2020openfermion,
  title={OpenFermion: the electronic structure package for quantum computers},
  author={McClean, Jarrod R and others},
  journal={Quantum Science and Technology},
  volume={5},
  number={3},
  pages={034014},
  year={2020},
  publisher={IOP Publishing}
}

@misc{Openfermion_tutorials,
  author       = {{OpenFermion Developers}},
  title        = {OpenFermion Tutorials},
  year         = {2024},
  howpublished = {\url{https://quantumai.google/openfermion/tutorials}},
  note         = {Accessed: 2025-06-09, See specifically the Hubbard model tutorial.}
}

@book{Sachdev2011quantum,
  title={Quantum Phase Transitions},
  author={Sachdev, Subir},
  year={2011},
  publisher={Cambridge University Press},
  edition={2nd}
}

@article{Savary2016quantum,
  title={Quantum spin liquids: a review},
  author={Savary, Lucile and Balents, Leon},
  journal={Reports on Progress in Physics},
  volume={80},
  number={1},
  pages={016502},
  year={2016},
  publisher={IOP Publishing}
}

@article{Shaydulin2023evidence,
  doi={10.48550/arXiv.2308.02341},
  title={Evidence of scaling advantage for the quantum approximate optimization algorithm on a classically intractable problem},
  author={Shaydulin, Ruslan and others},
  journal={arXiv preprint arXiv:2308.02341},
  year={2023}
}

@book{Mattis2006theory,
  title={The Theory of Magnetism Made Simple: An Introduction to Physical Concepts and to Some Useful Mathematical Methods},
  author={Mattis, Daniel C.},
  year={2006},
  publisher={World Scientific}
}

@article{Amico2008entanglement,
  doi={10.1103/RevModPhys.80.517},
  title={Entanglement in many-body systems},
  author={Amico, Luigi and Fazio, Rosario and Osterloh, Andreas and Vedral, Vlatko},
  journal={Reviews of Modern Physics},
  volume={80},
  number={2},
  pages={517},
  year={2008},
  publisher={APS}
}

@article{Hermanns2018physics,
  title={Physics of the Kitaev model: Fractionalization, emergent gauge fields, and topological order},
  author={Hermanns, Maria and Trebst, Simon and Rosch, Achim},
  journal={Annual Review of Condensed Matter Physics},
  volume={9},
  pages={17--33},
  year={2018},
  publisher={Annual Reviews}
}

@article{Sachdev1993gapless,
  doi={10.1103/PhysRevLett.70.3339},
  title={Gapless spin-fluid ground state in a random quantum Heisenberg magnet},
  author={Sachdev, Subir and Ye, Jinwu},
  journal={Physical Review Letters},
  volume={70},
  number={21},
  pages={3339},
  year={1993},
  publisher={APS}
}

@misc{Kitaev2015simple,
  author       = {Kitaev, Alexei},
  title        = {A simple model of quantum holography},
  year         = {2015},
  howpublished = {Talks at KITP, April 7 and May 27, 2015},
  note         = {Available at: \url{http://online.kitp.ucsb.edu/online/entangled15/}}
}

@article{Rosenhaus2019introduction,
  title={An introduction to the SYK model},
  author={Rosenhaus, Vladimir},
  journal={Journal of Physics A: Mathematical and Theoretical},
  volume={52},
  number={32},
  pages={323001},
  year={2019},
  publisher={IOP Publishing}
}

@article{Sachdev2017holographic,
  doi={10.1103/PhysRevLett.105.151602},
  title={Holographic metals and the fractionalized Fermi liquid},
  author={Sachdev, Subir},
  journal={Physical Review Letters},
  volume={105},
  number={15},
  pages={151602},
  year={2010},
  publisher={APS}
}

@article{Garcia2021digital,
  doi={10.1103/PhysRevD.109.105002},
  title={Sachdev-Ye-Kitaev model on a noisy quantum computer},
  author={Asaduzzaman, Muhammad and Jha, Raghav G and Sambasivam, Bharath},
  journal={Physical Review D},
  volume={109},
  number={10},
  pages={105002},
  year={2024},
  publisher={APS}
}

@article{Kitaev2003fault,
  doi={10.1016/S0003-4916(02)00018-0},
  title={Fault-tolerant quantum computation by anyons},
  author={Kitaev, Alexei Yu},
  journal={Annals of Physics},
  volume={303},
  number={1},
  pages={2--30},
  year={2003},
  publisher={Elsevier}
}

@article{Azad2022quantum,
  doi={10.1103/PhysRevA.106.012423},
 title = {Efficient quantum readout-error mitigation for sparse measurement outcomes of near-term quantum devices},
  author = {Yang, Bo and Raymond, Rudy and Uno, Shumpei},
  journal = {Phys. Rev. A},
  volume = {106},
  issue = {1},
  pages = {012423},
  numpages = {14},
  year = {2022},
  month = {Jul},
  publisher = {American Physical Society}
}

@article{Heyl2013dynamical,
  doi={10.1103/PhysRevLett.110.135704},
  title={Dynamical quantum phase transitions in the transverse-field Ising model},
  author={Heyl, Markus and Polkovnikov, Anatoli and Kehrein, Stefan},
  journal={Physical Review Letters},
  volume={110},
  number={13},
  pages={135704},
  year={2013},
  publisher={APS}
}

@article{Heyl2018dynamical,
  title={Dynamical quantum phase transitions: a review},
  author={Heyl, Markus},
  journal={Reports on Progress in Physics},
  volume={81},
  number={5},
  pages={054001},
  year={2018},
  publisher={IOP Publishing}
}

@article{Orus2014practical,
  doi={10.1016/j.aop.2014.06.013},
  title={A practical introduction to tensor networks: Matrix product states and projected entangled pair states},
  author={Or{\'u}s, Rom{\'a}n},
  journal={Annals of Physics},
  volume={349},
  pages={117--158},
  year={2014},
  publisher={Elsevier}
}

@article{Schollwock2011density,
  doi={10.1016/j.aop.2010.09.012},
  title={The density-matrix renormalization group in the age of matrix product states},
  author={Schollw{\"o}ck, Ulrich},
  journal={Annals of Physics},
  volume={326},
  number={1},
  pages={96--192},
  year={2011},
  publisher={Elsevier}
}

@article{Eisert2010colloquium,
  doi={10.1103/RevModPhys.82.277},
  title={Colloquium: Area laws for the entanglement entropy},
  author={Eisert, Jens and Cramer, Marcus and Plenio, Martin B.},
  journal={Reviews of Modern Physics},
  volume={82},
  number={1},
  pages={277},
  year={2010},
  publisher={APS}
}

@article{Yoshioka2024hunting,
  title={Hunting for quantum-classical crossover in condensed matter problems},
  author={Yoshioka, Nobuyuki and Okubo, Tsuyoshi and Suzuki, Yasunari and Koizumi, Yuki and Mizukami, Wataru},
  journal={npj Quantum Information},
  volume={10},
  number={1},
  pages={45},
  year={2024},
  publisher={Nature Publishing Group UK London}
}

@article{Vidal2008class,
  doi={10.1103/PhysRevLett.101.110501},
  title={A class of quantum many-body states that can be efficiently simulated},
  author={Vidal, Guifre},
  journal={Physical Review Letters},
  volume={101},
  number={11},
  pages={110501},
  year={2008},
  publisher={APS}
}

@article{Kim2017robust,
  doi={10.1103/PhysRevLett.119.140502},
  title={Robust entanglement renormalization on a noisy quantum computer},
  author={Kim, Isaac H. and Evenbly, Glen},
  journal={Physical Review Letters},
  volume={119},
  number={14},
  pages={140502},
  year={2017},
  publisher={APS}
}

@article{Evenbly2019representation,
  doi={10.1103/PhysRevA.99.012321},
  title={Representation of the multiscale entanglement renormalization ansatz as a quantum circuit},
  author={Evenbly, Glen},
  journal={Physical Review A},
  volume={99},
  number={1},
  pages={012321},
  year={2019},
  publisher={APS}
}

@article{Berezutskii2025tensor,
  doi={10.48550/arXiv.2503.08626},
  title={Tensor networks for quantum computing},
  author={Berezutskii, Aleksandr and Acharya, Atithi and Ellerbrock, Roman and Gray, Johnnie and Haghshenas, Reza and He, Zichang and Khan, Abid and Kuzmin, Viacheslav and Liu, Minzhao and Lyakh, Dmitry and others},
  journal={arXiv preprint arXiv:2503.08626},
  year={2025}
}

@article{Pan2022solving,
  doi={10.1103/PhysRevLett.129.090502},
  title={Solving the sampling problem of the Sycamore quantum circuits},
  author={Pan, Feng and Chen, Keyang and Pan, Pan},
  journal={Physical Review Letters},
  volume={129},
  number={9},
  pages={090502},
  year={2022},
  publisher={APS}
}

@article{Daley2022,
  title={Practical quantum advantage in quantum simulation},
  author={Daley, Andrew J. and others},
  journal={Nature},
  volume={607},
  number={7920},
  pages={667--676},
  year={2022},
  publisher={Nature Publishing Group}
}

@article{Bauer2020quantum,
  title={Quantum algorithms for quantum chemistry and quantum materials science},
  author={Bauer, Bela and Bravyi, Sergey and Motta, Mario and Chan, Garnet Kin-Lic},
  journal={Chemical Reviews},
  volume={120},
  number={22},
  pages={12685--12717},
  year={2020},
  publisher={ACS Publications}
}

@article{Tubman2020warm,
  doi={10.48550/arXiv.2009.07623},
  title={Warm-starting quantum optimization},
  author={Tubman, Norm M. and others},
  journal={arXiv preprint arXiv:2009.07623},
  year={2020}
}

@book{Fehske2007computational,
  title={Computational Many-Particle Physics},
  author={Fehske, Holger and Schneider, Ralf},
  year={2007},
  publisher={Springer}
}

@book{Martin2020electronic,
  title={Electronic Structure: Basic Theory and Practical Methods},
  author={Martin, Richard M. and Reining, Lucia and Ceperley, David M.},
  year={2020},
  publisher={Cambridge University Press}
}

@article{Sandvik2010computational,
  title={Computational studies of quantum spin systems},
  author={Sandvik, Anders W.},
  journal={AIP Conference Proceedings},
  volume={1297},
  number={1},
  pages={135--338},
  year={2010},
  organization={American Institute of Physics}
}

@article{Jones2015density,
  doi={10.1103/RevModPhys.87.897},
  title={Density functional theory: Its origins, rise to prominence, and future},
  author={Jones, R. O.},
  journal={Reviews of Modern Physics},
  volume={87},
  number={3},
  pages={897},
  year={2015},
  publisher={APS}
}

@article{Georges1996dynamical,
  doi={10.1103/RevModPhys.68.13},
  title={Dynamical mean-field theory of strongly correlated fermion systems and the limit of infinite dimensions},
  author={Georges, Antoine and Kotliar, Gabriel and Krauth, Werner and Rozenberg, Marcelo J.},
  journal={Reviews of Modern Physics},
  volume={68},
  number={1},
  pages={13},
  year={1996},
  publisher={APS}
}

@article{Hasnip2014density,
  title={Density functional theory in the solid state},
  author={Hasnip, Philip J. and Refson, Keith and Probert, Matt I. J. and Yates, Jonathan R. and Clark, Stewart J. and Pickard, Chris J.},
  journal={Philosophical Transactions of the Royal Society A: Mathematical, Physical and Engineering Sciences},
  volume={372},
  number={2011},
  pages={20130270},
  year={2014},
  publisher={The Royal Society}
}

@article{Troyer2005computational,
  doi={10.1103/PhysRevLett.94.170201},
  title={Computational studies of quantum spin systems},
  author={Troyer, Matthias and Wiese, Uwe-Jens},
  journal={Physical Review Letters},
  volume={94},
  number={17},
  pages={170201},
  year={2005},
  publisher={APS}
}

@article{Anisimov1991band,
  doi={10.1103/PhysRevB.44.943},
  title={Band theory and Mott insulators: Hubbard U instead of Stoner I},
  author={Anisimov, V. I. and Zaanen, Jan and Andersen, Ole K.},
  journal={Physical Review B},
  volume={44},
  number={3},
  pages={943},
  year={1991},
  publisher={APS}
}

@article{Yu2025clifford,
  doi={10.48550/arXiv.2502.11933},
  title={Clifford circuit based heuristic optimization of fermion-to-qubit mappings},
  author={Yu, Jeffery and Liu, Yuan and Sugiura, Sho and Van Voorhis, Troy and Zeytino{\u{g}}lu, Sina},
  journal={arXiv preprint arXiv:2502.11933},
  year={2025}
}

@article{Seeley2012bravyi,
  title={The Bravyi-Kitaev transformation for quantum computation of electronic structure},
  author={Seeley, Jacob T. and Richard, Martin J. and Love, Peter J.},
  journal={The Journal of chemical physics},
  volume={137},
  number={22},
  pages={224109},
  year={2012},
  publisher={AIP}
}

@article{Jiang2020optimal,
  doi={10.1103/PhysRevApplied.14.044022},
  title={Optimal fermion-to-qubit mapping in two spatial dimensions},
  author={Jiang, Zhang and Sung, Kevin J. and Kechedzhi, Kostyantyn and Smelyanskiy, Vadim N. and Boixo, Sergio},
  journal={Physical Review Applied},
  volume={14},
  number={4},
  pages={044022},
  year={2020},
  publisher={APS}
}

@article{Rubin2021fermionic,
  title={Fermionic quantum emulator},
  author={Rubin, Nicholas C. and others},
  journal={Quantum},
  volume={5},
  pages={562},
  year={2021},
  publisher={Verein zur F{\"o}rderung des Open Access Publizierens in den Naturwissenschaften}
}

@article{Qiskit_algorithms2023,
  title={{Qiskit Algorithms}: A library of quantum algorithms in Qiskit},
  author={{Qiskit Algorithms Team}},
  journal={Journal of Open Source Software},
  volume={8},
  number={88},
  pages={5349},
  year={2023},
  publisher={The Open Journal}
}

@book{Qiskit_textbook,
  author    = {{Qiskit Development Team}},
  title     = {Learn Quantum Computation using Qiskit},
  howpublished = {\url{https://qiskit.org/learn}},
  year      = {2024},
  note = {Contains detailed chapters and tutorials on algorithms like QPE and applications in chemistry and physics, which form the basis for the cited examples.}
}

@misc{cirq_developers_2024,
  author       = {{Cirq Developers}},
  title        = {Cirq},
  year         = {2024},
  publisher    = {Google},
  howpublished = {\url{https://quantumai.google/cirq}}
}

@article{Bergholm2022pennylane,
  doi={10.48550/arXiv.1811.04968},
  title={Pennylane: Automatic differentiation of hybrid quantum-classical computations},
  author={Bergholm, Ville and others},
  journal={arXiv preprint arXiv:1811.04968},
  year={2022}
}

@misc{Pennylane_documentation,
  author       = {{Xanadu}},
  title        = {PennyLane Documentation},
  year         = {2025},
  howpublished = {\url{https://docs.pennylane.ai/en/stable/}},
  note         = {Accessed: June 9, 2025}
}

@misc{Pennylane_qchem_module,
  author       = {{Xanadu}},
  title        = {PennyLane-QChem: Quantum Chemistry with PennyLane},
  year         = {2025},
  howpublished = {\url{https://docs.pennylane.ai/projects/qchem/en/latest/}},
  note         = {Accessed: June 9, 2025. This documentation covers the tools for building molecular Hamiltonians and performing quantum chemistry simulations applicable to materials science.}
}

@misc{microsoft_qsharp_overview,
  author       = {{Microsoft}},
  title        = {Introduction to the Quantum Programming Language Q\#},
  year         = {2025},
  howpublished = {\url{https://learn.microsoft.com/en-us/azure/quantum/qsharp-overview}},
  note         = {Accessed: June 9, 2025. Provides a high-level overview of the Q language and its features.}
}

@misc{microsoft_qdk_preview_blog,
  author       = {{Microsoft Quantum Team}},
  title        = {Introducing the Azure Quantum Development Kit Preview},
  year         = {2023},
  howpublished = {\url{https://quantum.microsoft.com/en-us/insights/blogs/qir/introducing-the-azure-quantum-development-kit-preview}},
  note         = {Blog post detailing the new, Rust-based QDK and its improvements.}
}

@article{Motta2020determining,
  title        = {Determining eigenstates and thermal states on a quantum computer using quantum imaginary time evolution},
  author       = {Motta, Mario and others},
  journal      = {Nature Physics},
  volume       = {16},
  number       = {2},
  pages        = {205--210},
  year         = {2020},
  publisher    = {Nature Publishing Group},
  note         = {This paper introduces the Quantum Imaginary Time Evolution (QITE) algorithm, a method highly relevant for finding ground states of strongly correlated systems targeted by Q.}
}

@article{cai2023quantum,
  doi={10.1103/RevModPhys.95.045005},
  title={Quantum error mitigation},
  author={Cai, Z. and others},
  journal={Reviews of Modern Physics},
  volume={95},
  number={4},
  pages={045005},
  year={2023},
  publisher={APS}
}

@article{Endo2021hybrid,
  doi={10.7566/JPSJ.90.032001},
  title={Hybrid quantum-classical algorithms and quantum error mitigation},
  author={Endo, Suguru and Cai, Zain and Benjamin, Simon C. and Yuan, Xiao},
  journal={Journal of the Physical Society of Japan},
  volume={90},
  number={3},
  pages={032001},
  year={2021},
  publisher={The Physical Society of Japan}
}

@article{Terhal2015quantum,
  doi={10.1103/RevModPhys.87.307},
  title={Quantum error correction for quantum memories},
  author={Terhal, Barbara M.},
  journal={Reviews of Modern Physics},
  volume={87},
  number={2},
  pages={307},
  year={2015},
  publisher={APS}
}

@article{laflamme_et_al_2024_benchmarking,
  doi={10.48550/arXiv.2409.08844},
    author = {Laflamme-SNIDER, Felix and others},
    title = {Benchmarking the performance of quantum computing software},
    journal = {arXiv preprint arXiv:2409.08844},
    year = {2024}
}

@article{Harrow2017quantum,
  doi={10.1038/nature23458},
  title={Quantum computational supremacy},
  author={Harrow, Aram W. and Montanaro, Ashley},
  journal={Nature},
  volume={549},
  number={7671},
  pages={203--209},
  year={2017},
  publisher={Nature Publishing Group}
}

@article{Carolan2020verifying,
  doi={10.1038/s41567-019-0747-6},
  title={Variational quantum unsampling on a quantum photonic processor},
  author={Carolan, Jacques and others},
  journal={Nature Physics},
  volume={16},
  number={3},
  pages={322--327},
  year={2020},
  publisher={Nature Publishing Group}
}

@article{Eisert2020certification,
  doi={10.1038/s42254-020-0186-4},
  title={Quantum certification and benchmarking},
  author={Eisert, Jens and Hangleiter, Dominik and Walk, Nathan and Roth, Ingo and Markham, Damian and Parekh, Rhea and Chabaud, Ulysse and Kashefi, Elham},
  journal={Nature Reviews Physics},
  volume={2},
  number={7},
  pages={382--390},
  year={2020},
  publisher={Nature Publishing Group}
}

@misc{qctrl_fire_opal,
  author       = {{Q-CTRL}},
  title        = {Fire Opal: Automate and optimize quantum algorithm execution},
  year         = {2025},
  howpublished = {\url{https://q-ctrl.com/fire-opal}},
  note         = {Accessed: June 9, 2025}
}

@article{de_lima_de_farias_2024_qpe,
  doi={10.48550/arXiv.2410.21517},
  title={Quantum Phase Estimation without Controlled Unitaries},
  author={de Lima de Farias, C. and others},
  journal={arXiv preprint arXiv:2410.21517},
  year={2024},
  note={This work explicitly demonstrates using classical phase retrieval algorithms to perform control-free QPE.}
}

@article{Miessen2023quantum,
  doi={10.1038/s43588-022-00374-2},
  title={Quantum algorithms for quantum dynamics},
  author={Miessen, Alexander and Ollitrault, Pauline J. and Tacchino, Francesco and Tavernelli, Ivano},
  journal={Nature Computational Science},
  volume={3},
  number={1},
  pages={25--37},
  year={2023},
  publisher={Nature Publishing Group}
}

@misc{bnl_novel_algorithms,
  author       = {{Brookhaven National Laboratory}},
  title        = {Quantum Computing Group | Novel Quantum Algorithms},
  year         = {2024},
  howpublished = {https://www.bnl.gov/compsci/quantum/novel-algorithms.php},
  note         = {Accessed: June 9, 2025. This project description outlines a research program to discover new quantum primitives from classical numerical transforms like Hermite and orthogonal polynomial transforms for fast-forwarding quantum simulations.}
}

@article{Murali2019noise,
  doi={10.48550/arXiv.2503.14592},
  title={Optimal and efficient qubit routing for quantum simulation},
  author={Kattem{\"o}lle, Joris and Burkard, Guido},
  journal={arXiv preprint arXiv:2503.14592},
  year={2025}
}

@article{shehata2025building,
  doi={10.48550/arXiv.2503.01787},
  title={Building a Software Stack for Quantum-HPC Integration},
  author={Shehata, Amir and others},
  journal={arXiv preprint arXiv:2503.01787},
  year={2025}
}

@article{mahesh2025conqure,
  doi={10.48550/arXiv.2505.02241},
  title={{CONQURE}: A Co-Execution Environment for Quantum and Classical Resources},
  author={Mahesh, Atulya and Mittal, Swastik and Mueller, Frank},
  journal={arXiv preprint arXiv:2505.02241},
  year={2025}
}

@misc{pasqal_standardization_2024,
    author = {Lefebvre, Catherine and Karagiannis, Konstantinos},
    title = {Quantum Computing Standardization},
    howpublished = {Protiviti Podcast},
    year = {2024},
    month = {May},
    note = {Transcript available at: \url{https://www.protiviti.com/us-en/podcast-transcript/quantum-computing-standardization}}
}

@article{Beverland2024early,
  doi={10.1103/PRXQuantum.5.020101},
    title = {Early Fault-Tolerant Quantum Computing},
    author = {Beverland, Michael E. and others},
    journal = {PRX Quantum},
    volume = {5},
    issue = {2},
    pages = {020101},
    year = {2024},
    publisher = {American Physical Society}
}

@article{Altman2021quantum,
  doi={10.1103/PRXQuantum.2.017003},
  title={Quantum simulators: Architectures and opportunities},
  author={Altman, Ehud and others},
  journal={PRX Quantum},
  volume={2},
  number={1},
  pages={017003},
  year={2021},
  publisher={APS}
}

@article{Brennan2022optimizing,
  doi={10.1103/PhysRevA.105.052431},
  title={Optimizing and compiling weighted-graph-based algorithms for the quantum approximate optimization algorithm on neutral-atom quantum computers},
  author={Brennan, B. D. and others},
  journal={Physical Review A},
  volume={105},
  number={5},
  pages={052431},
  year={2022},
  publisher={APS},
  note={This work focuses on compilation strategies that exploit the unique features of neutral-atom hardware.}
}

@article{Krinner2022superconducting,
    doi={10.1038/s41586-022-04566-8},
    title={Realizing repeated quantum error correction in a distance-three surface code},
    author={Krinner, S. and others},
    journal={Nature},
    volume={605},
    pages={669--674},
    year={2022}
}

@article{Graham2022multi,
    doi={10.1038/s41586-022-04603-6},
    title={Multi-qubit entanglement and algorithms on a neutral-atom quantum computer},
    author={Graham, T. M. and others},
    journal={Nature},
    volume={604},
    pages={457--462},
    year={2022}
}

@misc{ibm_ftqc_roadmap_2025,
  author       = {{IBM Research}},
  title        = {IBM lays out clear path to fault-tolerant quantum computing},
  year         = {2025},
  month        = {June},
  howpublished = {\url{https://www.ibm.com/quantum/blog/large-scale-ftqc}},
  note         = {Accessed: June 10, 2025. This blog post details IBM's roadmap towards 1000s of qubits and discusses the software and connectivity improvements (e.g., Nighthawk's square lattice topology) needed to make them effective.}
}

@misc{the_quantum_insider_roadmaps_2025,
  author       = {{The Quantum Insider}},
  title        = {Quantum Computing Roadmaps: A Look at The Maps And Predictions of Major Quantum Players},
  year         = {2025},
  month        = {May},
  howpublished = {\url{https://thequantuminsider.com/2025/05/16/quantum-computing-roadmaps-a-look-at-the-maps-and-predictions-of-major-quantum-players/}},
  note         = {Accessed: June 10, 2025. This article aggregates roadmaps from multiple companies (Pasqal, IonQ, etc.), all pointing to larger qubit counts and the corresponding need for software to enable practical applications.}
}

@misc{Das2024quantum,
    author       = {{IBM Quantum}},
    title        = {Quantum-centric supercomputing: The next wave of computing},
    year         = {2022},
    howpublished = {\url{https://www.ibm.com/quantum/blog/next-wave-quantum-centric-supercomputing}},
    note         = {IBM Quantum blog. Outlines the vision for integrating large-scale QPUs with classical supercomputers, which demands sophisticated software to manage the increased resource pool.}
}

@article{Smejkal2022altermagnetism,
  doi={10.1103/PhysRevX.12.040501},
  title={Emerging research landscape of altermagnetism},
  author={{\v{S}}mejkal, Libor and Sinova, Jairo and Jungwirth, Tom{\'a}{\v{s}}},
  journal={Physical Review X},
  volume={12},
  number={4},
  pages={040501},
  year={2022},
  publisher={APS}
}

@article{Smejkal2022emerging,
  doi={10.1038/s41578-022-00430-3},
  title={Anomalous Hall antiferromagnets},
  author={{\v{S}}mejkal, Libor and MacDonald, A. H. and Sinova, Jairo and Nakatsuji, Satoru and Jungwirth, Tom{\'a}{\v{s}}},
  journal={Nature Reviews Materials},
  volume={7},
  number={6},
  pages={482--496},
  year={2022},
  publisher={Nature Publishing Group}
}

@article{Roy2019topological,
  doi={10.1103/PhysRevB.99.214505},
  title={Topological superconductivity in a weyl-semimetal-superconductor heterostructure},
  author={Roy, Bitan and Goswami, Pallab},
  journal={Physical Review B},
  volume={99},
  number={21},
  pages={214505},
  year={2019},
  publisher={APS}
}

@article{Paudyal2025phonon,
  doi={10.48550/arXiv.2506.03250},
  title={Phonon-Mediated Intrinsic Topological Superconductivity in Fermi Arcs},
  author={Paudyal, Anup and others},
  journal={arXiv preprint arXiv:2506.03250},
  year={2025}
}

@article{Mcnamara2025novel,
  doi={10.48550/arXiv.2506.03200},
  title={Novel Experimental Platform to realize One-dimensional Quantum Fluids},
  author={McNamara, Stephanie and Parajuli, Prabin and Paul, Sutirtha and Warren, Garfield and Del Maestro, Adrian and Sokol, Paul E.},
  journal={arXiv preprint arXiv:2506.03200},
  year={2025}
}

@article{Sokol2025experimental,
  doi={10.1007/s10909-025-03295-2},
  title={Experimental Realization of One-Dimensional Helium},
  author={Sokol, P. E. and others},
  journal={Journal of Low Temperature Physics},
  volume={220},
  pages={124--135},
  year={2025},
  publisher={Springer}
}

@article{Boschini2024time,
  doi={10.1103/RevModPhys.96.015003},
  title={Time-resolved ARPES studies of quantum materials},
  author={Boschini, F. and Zonno, M. and Damascelli, A. and others},
  journal={Reviews of Modern Physics},
  volume={96},
  number={1},
  pages={015003},
  year={2024},
  publisher={APS}
}

@article{Ulstrup2023probing,
  title={Probing electronic structure and ultrafast dynamics in quantum materials with time- and angle-resolved photoemission spectroscopy},
  author={Ulstrup, S. and others},
  journal={Journal of Physics: Condensed Matter},
  volume={35},
  number={39},
  pages={393001},
  year={2023},
  publisher={IOP Publishing}
}

@article{Dai2025probing,
  doi={10.48550/arXiv.2501.05068},
  title={Probing Non-Equilibrium Pair-Breaking and Quasiparticle Dynamics in Nb Superconducting Resonators Under Magnetic Fields},
  author={Dai, X. and others},
  journal={arXiv preprint arXiv:2501.05068},
  year={2025}
}

@article{Deutsch2018eigenstate,
  doi={10.1088/1361-6633/aac9f1},
  title={Eigenstate thermalization hypothesis},
  author={Deutsch, Joshua M},
  journal={Reports on Progress in Physics},
  volume={81},
  number={8},
  pages={082001},
  year={2018},
  publisher={IOP Publishing}
}

@article{Dalessio2016from,
  doi={10.1080/00018732.2016.1198134},
  title={From quantum chaos and eigenstate thermalization to statistical mechanics and thermodynamics},
  author={D'Alessio, Luca and Kafri, Yariv and Polkovnikov, Anatoli and Rigol, Marcos},
  journal={Advances in Physics},
  volume={65},
  number={3},
  pages={239--362},
  year={2016},
  publisher={Taylor \& Francis}
}

@article{Bhattacharya2025thermalization,
  doi={10.1103/PhysRevB.111.195153},
  title = {Thermalization of a closed Sachdev-Ye-Kitaev system in the thermodynamic limit},
  author = {Jaramillo, Santiago Salazar and Jha, Rishabh and Kehrein, Stefan},
  journal = {Phys. Rev. B},
  volume = {111},
  issue = {19},
  pages = {195153},
  numpages = {28},
  year = {2025},
  month = {May},
  publisher = {American Physical Society}
}

@article{nayak2008non,
  doi={10.1103/RevModPhys.80.1083},
  title={Non-Abelian anyons and topological quantum computation},
  author={Nayak, Chetan and Simon, Steven H and Stern, Ady and Freedman, Michael and Das Sarma, Sankar},
  journal={Reviews of Modern Physics},
  volume={80},
  number={3},
  pages={1083--1159},
  year={2008},
  publisher={APS}
}

@article{GoogleQuantumAI2025qec,
  doi={10.1038/s41586-024-08449-y},
  title={Quantum error correction below the surface code threshold},
  author={{Google Quantum AI and Collaborators}},
  journal={Nature},
  volume={638},
  number={8052},
  pages={920--926},
  year={2025},
  publisher={Nature Publishing Group}
}

@article{Bluvstein2024logical,
  doi={10.1038/s41586-023-06927-3},
  title={Logical quantum processor based on reconfigurable atom arrays},
  author={Bluvstein, Dolev and Evered, Simon J. and Geim, Alexandra A. and Li, Sophie H. and Zhou, Hengyun and Manovitz, Tom and Ebadi, Sepehr and Cain, Madelyn and Kalinowski, Marcin and Hangleiter, Dominik and others},
  journal={Nature},
  volume={626},
  number={7997},
  pages={58--65},
  year={2024},
  publisher={Nature Publishing Group}
}

@article{Kim2023utility,
  doi={10.1038/s41586-023-06096-3},
  title={Evidence for the utility of quantum computing before fault tolerance},
  author={Kim, Youngseok and Eddins, Andrew and Anand, Sajant and Wei, Ken Xuan and van den Berg, Ewout and Rosenblatt, Sami and Nayfeh, Hasan and Wu, Yantao and Zaletel, Michael and Temme, Kristan and Kandala, Abhinav},
  journal={Nature},
  volume={618},
  number={7965},
  pages={500--505},
  year={2023},
  publisher={Nature Publishing Group}
}

@article{Begusic2024fast,
  doi={10.1126/sciadv.adk4321},
  title={Fast and converged classical simulations of evidence for the utility of quantum computing before fault tolerance},
  author={Begu{\v{s}}i{\'c}, Tomislav and Gray, Johnnie and Chan, Garnet Kin-Lic},
  journal={Science Advances},
  volume={10},
  number={3},
  pages={eadk4321},
  year={2024},
  publisher={American Association for the Advancement of Science}
}

@article{Tindall2024efficient,
  doi={10.1103/PRXQuantum.5.010308},
  title={Efficient tensor network simulation of {IBM}'s Eagle kicked {I}sing experiment},
  author={Tindall, Joseph and Fishman, Matthew and Stoudenmire, E. Miles and Sels, Dries},
  journal={PRX Quantum},
  volume={5},
  pages={010308},
  year={2024},
  publisher={American Physical Society}
}

@article{Shao2024antiferromagnetic,
  doi={10.1038/s41586-024-07689-2},
  title={Antiferromagnetic phase transition in a {3D} fermionic {H}ubbard model},
  author={Shao, Hou-Ji and others},
  journal={Nature},
  volume={632},
  number={8024},
  pages={267--272},
  year={2024},
  publisher={Nature Publishing Group}
}

\end{document}